\documentclass[10pt,twocolumn]{article}
\usepackage[top=1.85cm, bottom=1.85cm, left=1.7cm, right=1.7cm]{geometry}
\usepackage{times}
\usepackage[hyphens]{url}  %
\usepackage[keeplastbox]{flushend}  %
\usepackage{graphicx}  %
\usepackage{xcolor}
\usepackage{flushend}
\usepackage[sort&compress,numbers]{natbib}

\usepackage{booktabs}
\usepackage{graphicx}
\usepackage{paralist}
\usepackage{enumitem}
\usepackage{array}

\usepackage[small,bf]{caption}
\usepackage{caption}
\usepackage{subfigure}

\usepackage[hang,flushmargin]{footmisc}

\usepackage[bookmarks=true,%
            bookmarksnumbered=true,%
            colorlinks=true,%
            linkcolor=linkcol,%
            citecolor=citecol,%
            urlcolor=urlcol,%
            hypertexnames=true,%
            pdfpagelabels,
      ]%
      {hyperref}

\usepackage[OT2,OT1]{fontenc}
\newcommand\cyr
{
\renewcommand\rmdefault{wncyr}
\renewcommand\sfdefault{wncyss}
\renewcommand\encodingdefault{OT2}
\normalfont
\selectfont
}
\DeclareTextFontCommand{\textcyr}{\cyr}

\usepackage[compact]{titlesec}
\titlespacing*{\section}{0pt}{*2.5}{2.5pt} 
\titlespacing{\subsection}{0pt}{*2}{2pt}

\definecolor{linkcol}{rgb}{0,0,0.5}
\definecolor{citecol}{rgb}{0,0.5,0.3}
\definecolor{urlcol}{rgb}{0.3,0,0}

\renewenvironment{thebibliography}[1]{
  \begin{oldthebibliography}{#1}
    \setlength{\itemsep}{0.1em}
    \setlength{\parskip}{0.1em}
}
{
  \end{oldthebibliography}
}

\renewcommand{\footnoterule}{%
  \kern -3pt
  \hrule width 1in
  \kern 2pt
}

\usepackage{xspace}

\makeatletter
\def\url@leostyle{%
  \@ifundefined{selectfont}{\def\UrlFont{}}%
  {\def\UrlFont{}}%
}
\makeatother
\begin{document}
\title{\bf Disturbed YouTube for Kids: Characterizing and Detecting Inappropriate Videos Targeting Young Children\thanks{Published at the 14th International Conference on Web and Social Media (ICWSM 2020). Please cite the ICWSM version.}}

\author{Kostantinos Papadamou$^\star$, Antonis Papasavva$^\star$, Savvas Zannettou$^\ast$, Jeremy Blackburn$^\dagger$ \\
\large Nicolas Kourtellis$^\ddagger$, Ilias Leontiadis$^\ddagger$, Gianluca Stringhini$^\diamond$, Michael Sirivianos$^\star$ \\[0.5ex]
\normalsize $^\star$Cyprus University of Technology, $^\ast$Max-Planck-Institut f\"{u}r Informatik, $^\dagger$Binghamton University, \\
\normalsize $^\ddagger$Telefonica Research, $^\diamond$Boston University \\
\normalsize \{ck.papadamou,as.papasavva\}@edu.cut.ac.cy, szannett@mpi-inf.mpg.de, jblackbu@binghamton.edu \\%
\normalsize \{nicolas.kourtellis,ilias.leontiadis\}@telefonica.com, gian@bu.edu,
michael.sirivianos@cut.ac.cy}
\date{}

\maketitle

\begin{abstract}
A large number of the most-subscribed YouTube channels target children of very young age.
Hundreds of toddler-oriented channels on YouTube feature inoffensive, well produced, and educational videos.
Unfortunately, inappropriate content that targets this demographic is also common. 
YouTube's algorithmic recommendation system regrettably suggests inappropriate content because some of 
it mimics or is derived from otherwise appropriate content.
Considering the risk for early childhood development, and an increasing trend in toddler's consumption of YouTube media, this is a worrisome problem.
In this work, we build a classifier able to discern inappropriate content that targets toddlers on YouTube with $84.3\%$ accuracy, and leverage it to perform a large-scale, quantitative characterization that reveals some of the risks of YouTube media consumption by young children.
Our analysis reveals that YouTube is still plagued by such disturbing videos and its currently deployed counter-measures are ineffective in terms of detecting them in a timely manner.
Alarmingly, using our classifier we show that young children are not only able, but \emph{likely} to encounter disturbing videos when they randomly browse the platform starting from benign videos.
\end{abstract}

\section{Introduction} 
\label{sec:introduction}
\begin{figure}[t!]
\centering
\includegraphics[width=\columnwidth]{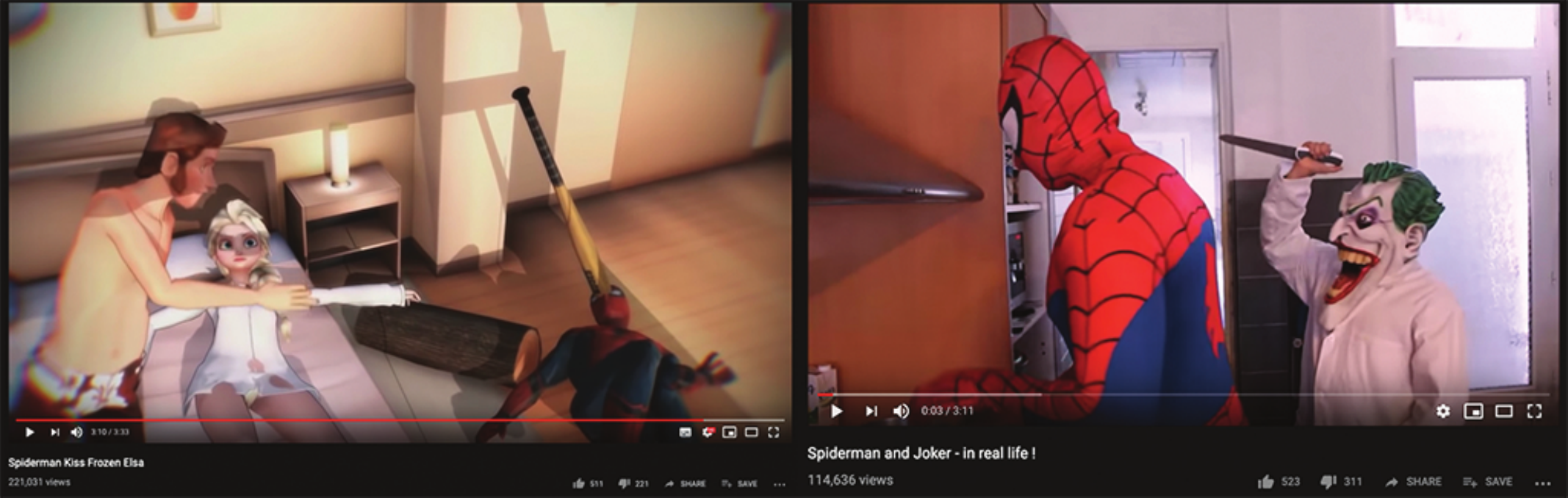}
\caption{Examples of disturbing videos, i.e. inappropriate videos that target toddlers.}
\label{fig:disturbing_video_example} 
\end{figure}

YouTube has emerged as an alternative to traditional children's TV, and a plethora of popular children's videos can be found on the platform. 
For example, consider the millions of subscribers that the most popular toddler-oriented YouTube channels have: ChuChu TV is the most-subscribed ``child-themed'' channel, with 19.9M subscribers~\cite{statista2018mostpopularchannels} as of September 2018. 
While most toddler-oriented content is inoffensive, and is actually entertaining or educational, recent reports have highlighted the trend of inappropriate content targeting this demographic~\cite{bbc2017disturbingyoutube,nytimes2017youtubekids}.
Borrowing the terminology from the early press articles on the topic, we refer to this new class of content as \emph{disturbing}. 
A prominent example of this trend is the Elsagate controversy~\cite{elsagate2017Reddit,theverge2017elsagate}, where malicious users uploaded videos featuring popular cartoon characters like Spiderman, Disney's Frozen, Mickey Mouse, etc., combined with disturbing content containing, for example, mild violence and sexual connotations. 
These disturbing videos usually include an innocent thumbnail aiming at tricking the toddlers and their custodians.
Fig.~\ref{fig:disturbing_video_example} shows examples of such videos.
The issue at hand is that these videos have hundreds of thousands of views, more likes than dislikes, and have been available on the platform since 2016.

In an attempt to offer a safer online experience for its young audience, YouTube launched the YouTube Kids application\footnote{\url{https://www.youtube.com/yt/kids/}}, which equips parents with several controls enabling them to decide what their children are allowed to watch on YouTube.
Unfortunately, despite YouTube's attempts to curb the phenomenon of inappropriate videos for toddlers, disturbing videos still appear, even in YouTube Kids~\cite{unsafeYoutubeForKids2018}, due to the difficulty of identifying them.
An explanation for this may be that YouTube relies heavily on users reporting videos they consider 
disturbing\footnote{\url{https://support.google.com/youtube/answer/2802027}}, and then YouTube employees manually inspecting them.
However, since the process involves manual labor, the whole mechanism does not easily scale to the amount of videos that a platform like YouTube serves.

In this paper, we provide the first study of toddler-oriented disturbing content on YouTube. 
For the purposes of this work, we extend the definition of a toddler to any child aged between 1 and 5 years. 
Our study comprises three phases.
First, we aim to characterize the phenomenon of inappropriate videos geared towards toddlers.
To this end, we collect, manually review, and characterize toddler-oriented videos (both Elsagate-related and other child-related videos), as well as random and popular videos.
For a more detailed analysis of the problem, we label these videos as one of four categories: 1)~suitable; 2)~disturbing; 3)~restricted (equivalent to 
MPAA's\footnote{MPAA stands for Motion Picture Association of America (MPAA) \url{https://www.mpaa.org/film-ratings/}} NC-17 and R categories); 
and 4)~irrelevant videos (see Section~\ref{subsec:manual-annotation}).
Our characterization confirms that unscrupulous and potentially profit-driven uploaders create disturbing videos with similar characteristics as benign 
toddler-oriented videos in an attempt to make them show up as recommendations to toddlers browsing the platform.

Second, we develop a deep learning classifier to automatically detect disturbing videos.
Even though this classifier performs better than baseline models, it still has a lower than desired performance.
In fact, this low performance reflects the high degree of similarity between disturbing and suitable videos or restricted videos that do not target toddlers.
It also reflects the subjectivity in deciding how to label these controversial videos, as confirmed by our trained annotators' experience.
For the sake of our analysis in the next steps, we collapse the initially defined labels into two categories and develop a more accurate classifier that is able to discern inappropriate from appropriate videos.
Our experimental evaluation shows that the developed binary classifier outperforms several baselines with an accuracy of 84.3\%.

In the last phase, we leverage the developed classifier to understand how prominent the problem at hand is.
From our analysis on different subsets of the collected videos, we find that 1.1\% of the 
233,337 Elsagate-related, and 0.5\% of the 154,957 other children-related collected videos
are inappropriate for toddlers, which indicates that the problem is not negligible.
To further assess how safe YouTube is for toddlers, we run a live simulation in which we mimic a toddler randomly clicking on YouTube's suggested videos.
We find that there is a 3.5\% chance that a toddler following YouTube's recommendations will encounter an inappropriate video within ten hops if she starts from a video that appears among the top ten results of a toddler-appropriate keyword search (e.g., Peppa Pig).

Last, our assessment on YouTube's current mitigations shows that the platform struggles to keep up with the problem: only 20.5\% and 2.5\% of our manually reviewed disturbing and restricted videos, respectively, have been removed by YouTube.

\noindent \textbf{Contributions.} In summary, our contribution is threefold:
\begin{compactenum}
	\item We undertake a large-scale analysis of the disturbing videos problem that is currently plaguing YouTube.
	\item We propose a reasonably accurate classifier that can be used to discern disturbing videos which target toddlers.
	\item We make publicly available the classifier\footnote{\url{https://tinyurl.com/disturbed-youtube-classifier}}, the manually reviewed ground truth dataset that consists of 4,797 videos, and the metadata of all the collected and examined videos\footnote{\url{https://zenodo.org/record/3632781}} so that the research community can build on our results to further investigate the problem.
\end{compactenum}

\section{Methodology} \label{sec:methodology}
\subsection{Data Collection}
For our data collection, we use the YouTube Data API\footnote{\url{https://developers.google.com/youtube/v3/}}, which provides metadata of videos 
uploaded on YouTube. 
Unfortunately, YouTube does not provide an API for retrieving videos from YouTube Kids.
We collect a set of seed videos using four different approaches. 
First, we use information from /r/ElsaGate, a subreddit dedicated to raising awareness about disturbing videos problem~\cite{elsagate2017Reddit}. 
Second, we use information from /r/fullcartoonsonyoutube, a subreddit dedicated to listing cartoon videos available on YouTube. 
The other two approaches focus on obtaining a set of random and popular videos.

Specifically:
1) we create a list of 64 keywords\footnote{\url{https://tinyurl.com/yxpf73j4}} by extracting n-grams from the title of videos posted on /r/ElsaGate.
Subsequently, for each keyword, we obtain the first 30 videos as returned by YouTube's Data API search functionality.
This approach resulted in the acquisition of 893 seed videos.
Additionally, we create a list of 33 channels\footnote{\url{https://tinyurl.com/y5zhy4vt}}, which are mentioned by users on /r/ElsaGate because of publishing 
disturbing videos~\cite{theverge2017elsagate,elsagate2017Reddit}.
Then, for each channel we collect all their videos, hence acquiring a set of 181 additional seed videos;
2) we create a list of 83 keywords\footnote{\url{https://tinyurl.com/y23xxl3c}} by extracting n-grams from the title of videos posted on /r/fullcartoonsonyoutube. 
Similar to the previous approach, for each keyword, we obtain the first 30 videos as returned by the YouTube's Data API search functionality, hence acquiring another 2,342 seed videos;
3) to obtain a random sample of videos, we a REST API\footnote{\url{https://randomyoutube.net/api}} that provides random YouTube video identifiers which we then download using the YouTube Data API. This approach resulted in the acquisition of 8,391 seed random videos; and
4) we also collect the most popular videos in the USA, the UK, Russia, India, and Canada, between November 18 and November 21, 2018, hence acquiring another 500 seed videos.

Using these approaches, we collect 12,097 unique seed videos.
However, this dataset is not big enough to study the idiosyncrasies of this problem.
Therefore, to expand our dataset, for each seed video we iteratively collect the top 10 recommended videos associated with it, as returned by the YouTube Data API, for up to three hops within YouTube’s recommendation graph.
We note that for each approach we use API keys generated from different accounts.
Table~\ref{tab:dataset_details} summarizes the collected dataset.
In total, our dataset comprises 12K seed videos and 844K videos that are recommended from the seed videos. 
Note, that there is a small overlap between the videos collected across the approaches, hence the number of total videos is slightly smaller than the sum of all videos for the four approaches. 

For each video in our dataset, we collect the following data descriptors: 1) title and description; 2) thumbnail; 3) tags; and 4) video statistics like number of views, likes, dislikes, etc.
We chose to use these four data collection approaches for three reasons:
1) to get more breadth into children's content on YouTube, instead of only collecting Elsagate-related videos; 2) to examine and analyze different types of videos while also assessing the degree of the disturbing videos problem in these types of videos; and 3) to train a classifier for detecting disturbing videos able to generalize to the different types of videos available on YouTube.

\noindent \textbf{Ethics.} 
For this study we only collect publicly available data, while making no attempt to de-anonymize users.
In addition, all the manual annotators are informed adults.

\begin{table}[t!]
\centering
\resizebox{0.7\columnwidth}{!}{%
\begin{tabular}{lrr}
\toprule
\textbf{Crawling Strategy} & \textbf{\#Seed} & \textbf{\#Recommended}\\
\midrule
Elsagate-related & 1,074 & 232,263 \\
Other Child-related	& 2,342	& 152,615 \\
Random & 8,391	& 473,516 \\
Popular & 500 	& 10,474 \\
\midrule
\textbf{Total} & 12,097 & 844,702 \\
\toprule
\end{tabular}}
\caption{Overview of the collected data: number of seed videos and number of their recommended videos acquired using each crawling strategy.}
\label{tab:dataset_details}
\end{table}

\subsection{Manual Annotation Process}\label{subsec:manual-annotation}
To get labeled data, we manually review a 5K videos subset of the collected dataset by inspecting their video content, title, thumbnail, and tags.
Each video is presented to three annotators that inspect its content and metadata to assign one of the following labels:

\noindent \textbf{Suitable:} 
A video is {\em suitable} when its content is appropriate for toddlers (aged 1-5 years) and it is relevant to their typical interests.
Some examples include normal cartoon videos, children's songs, and educational videos (e.g., learning colors). 
In other words, any video that can be classified as G by the MPAA and its target audience is toddlers.
    
\noindent \textbf{Disturbing:} 
A video is {\em disturbing} when it targets toddlers but it contains sexual hints, sexually explicit or abusive/inappropriate language, graphic nudity, child abuse (e.g., children hitting each other), scream and horror sound effects, scary scenes or characters (e.g., injections, attacks by insects, etc.).
In general, any video targeted at toddlers that should be classified as PG, PG-13, NC-17, or R by MPAA is considered disturbing.

\noindent \textbf{Restricted:} 
We consider a video {\em restricted} when it does not target toddlers and it contains content that is inappropriate for individuals under the age of 17 (rated as R or NC-17 according to MPAA).
Such videos usually contain sexually explicit language, graphic nudity, pornography, violence (e.g., gaming videos featuring violence, or life-like violence, etc.), abusive/inappropriate language, online gambling, drug use, alcohol, or upsetting situations and activities.

\noindent \textbf{Irrelevant:} 
We consider a video {\em irrelevant} when it contains appropriate content that is not relevant to a toddler's interests. 
That is, videos that are not disturbing or restricted but are only suitable for school-aged children (aged 6-11 years), adolescents (aged 12-17 years) and adults, like gaming videos (e.g., Minecraft) or music videos (e.g., a video clip of John Legend's song) reside in this category.
In general, G, PG and PG-13 videos that do not target toddlers are considered irrelevant.

We elect to use these labels for our annotation process instead of adopting the five MPAA ratings for two reasons.
First, our scope is videos that would normally be rated as PG, PG-13, R, and NC-17 but target very young audiences.
We consider such targeting a malevolent activity that needs to be treated separately.
At the same time, we have observed that a significant portion of videos that would normally be rated as R or NC-17 are already classified by YouTube as ``age-restricted'' and target either adolescents or adults.
Second, YouTube does not use MPAA ratings to flag videos, thus, a ground truth dataset with such labels is not available.

\noindent \textbf{Sampling Process.}
Our aim is to create a ground truth dataset that enables us to: 1) understand the main characteristics of disturbing toddler-oriented videos compared to suitable children videos on YouTube; and 2) train a deep learning model that will detect disturbing videos with an acceptable performance while being able to generalize to the various types of videos available on the platform.
To this end, we use the following videos for the annotation process.
1) We randomly select 1,000 of the 2,342 seed child-related videos aiming to get suitable videos.
2) Since the Elsagate-related collected videos are likely to include disturbing videos, we select all the seed Elsagate-related videos (1,074), as well as a small, randomly selected set (1,171) of their recommended videos.
3) To get a sample of restricted videos, we randomly select 500 of the 2,597 age-restricted videos in our dataset.
4) To ensure that we include irrelevant videos, we select all the seed popular videos (500) as well as a small set (1,000) of the 8,391 randomly collected videos.

\begin{table}[t!]
\centering
\resizebox{\columnwidth}{!}{%
\begin{tabular}{lrrrr}
\toprule
 & \textbf{\#Suitable} & \textbf{\#Disturbing} & \textbf{\#Restricted} & \textbf{\#Irrelevant} \\
\midrule
Elsagate-related & 805 & 857 & 324 & 394 \\
Other & 650 & 47 & 21 & 243 \\
Child-related & & & & \\
Random & 27 & 5 & 67 & 867 \\
Popular & 31 & 20 & 7 & 432 \\
\midrule
\textbf{Total} & 1,513 & 929 & 419 & 1,936 \\
\toprule
\end{tabular}%
}
\caption{Summary of our final ground truth dataset.}
\label{tab:final_groundtruth_dataset_details}
\end{table}

\noindent \textbf{Manual Annotation.}
The annotation process is carried out by two of the authors of this study and 76 undergraduate students (aged 20-24 years), both male and female.
Each video is annotated by the two authors and one of the undergraduate students.
The students come from different backgrounds and receive no specific training with regard to our study.
To ease the annotation process, we develop a platform that includes a clear description of the annotation task, our labels, as well as all the video information that an annotator needs in order to inspect and correctly annotate a video.

After obtaining all the annotations, we compute the Fleiss agreement score ($\kappa$)~\cite{fleiss1971measuring} across all annotators: we find $\kappa=0.60$, which is considered ``moderate'' agreement.
We also assess the level of agreement between the two authors, as we consider them experienced annotators, finding $\kappa=0.69$, which is considered ``substantial'' agreement. 
Finally, for each video we assign one of the labels according to the majority agreement of all the annotators, except a small percentage ($4\%$) where all annotators disagreed, which we also exclude from our ground truth dataset.
Table~\ref{tab:final_groundtruth_dataset_details} summarizes our ground truth dataset, which includes 1,513 suitable, 929 disturbing, 419 restricted, and 1,936 irrelevant videos.

\begin{figure*}[t!]
\centering
\subfigure[]{\includegraphics[width=0.23\textwidth, height=1.4in]{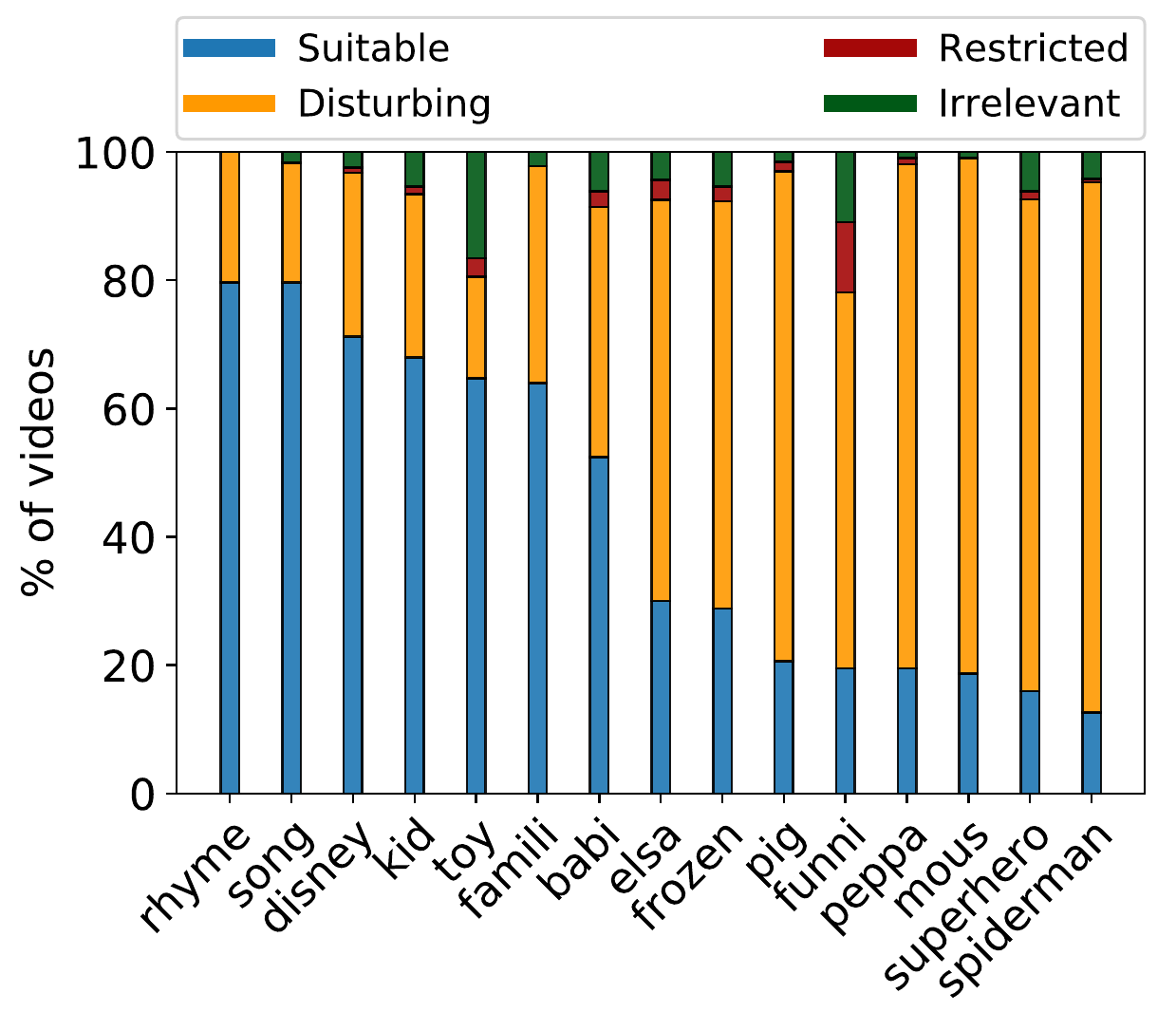}\label{fig:headline_normalized_mean_scores_elsagate}}
\subfigure[]{\includegraphics[width=0.23\textwidth, height=1.4in]{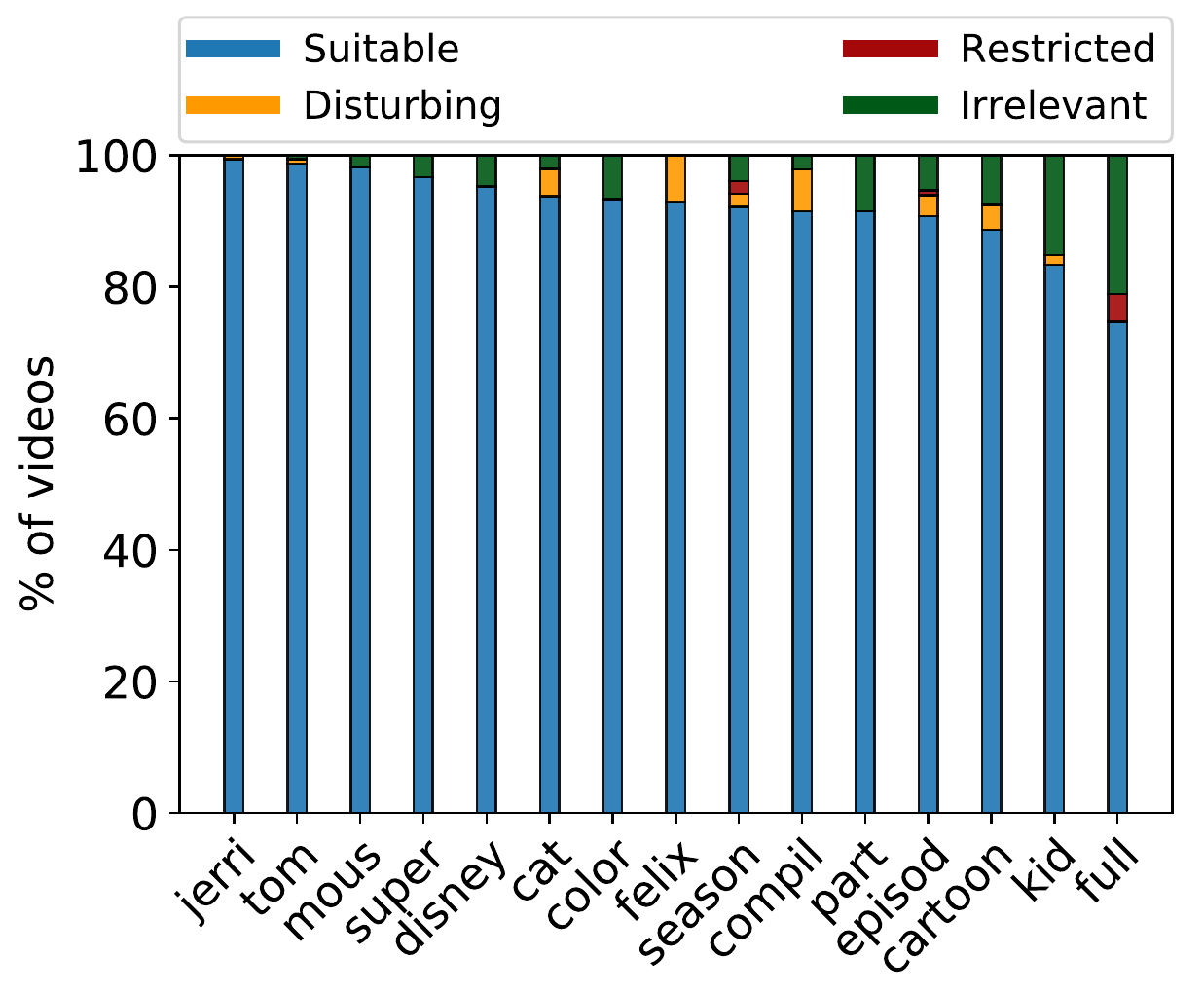}\label{fig:headline_normalized_mean_scores_childrelated}}
\subfigure[]{\includegraphics[width=0.23\textwidth, height=1.4in]{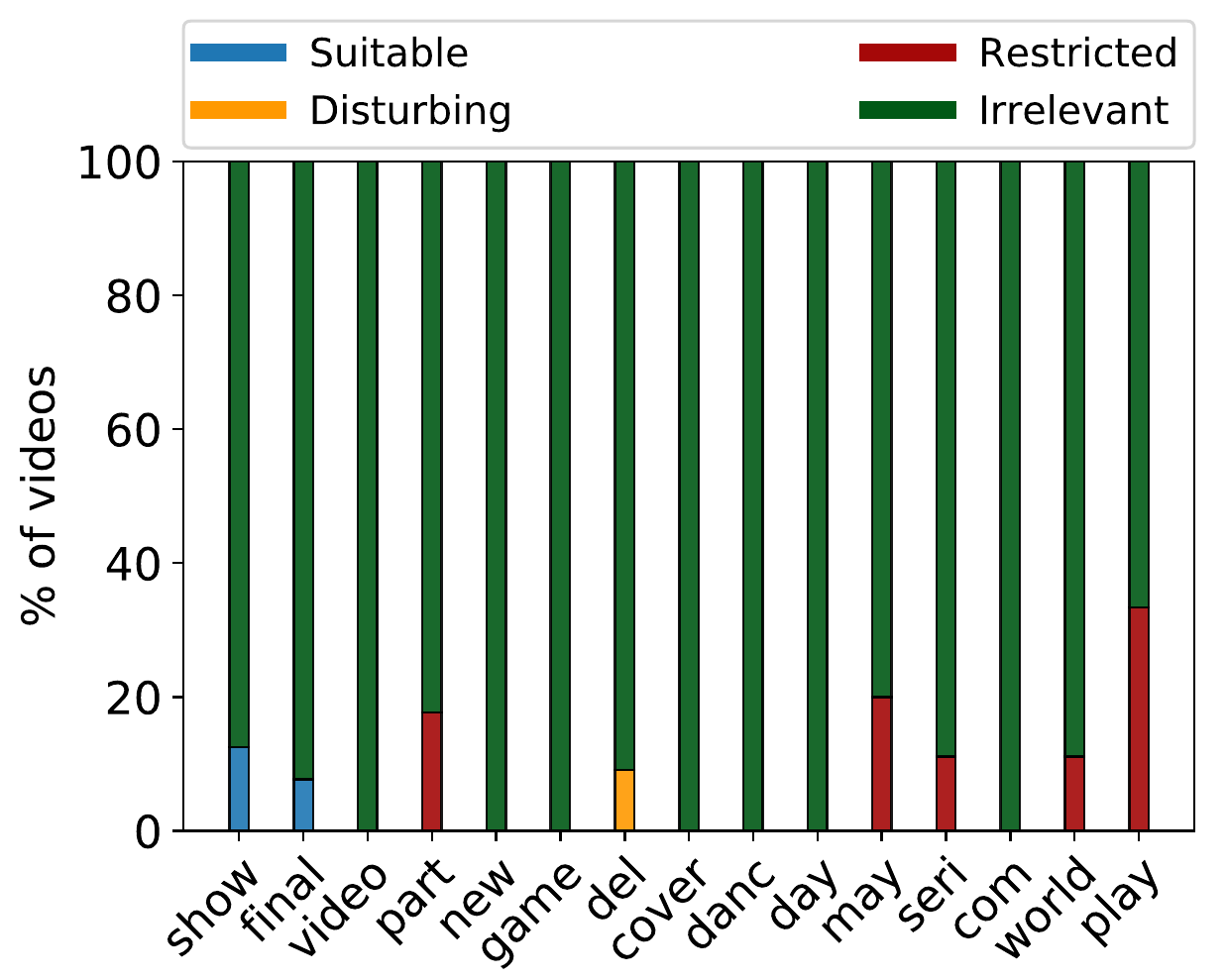}\label{fig:headline_normalized_mean_scores_random}}
\subfigure[]{\includegraphics[width=0.23\textwidth, height=1.4in]{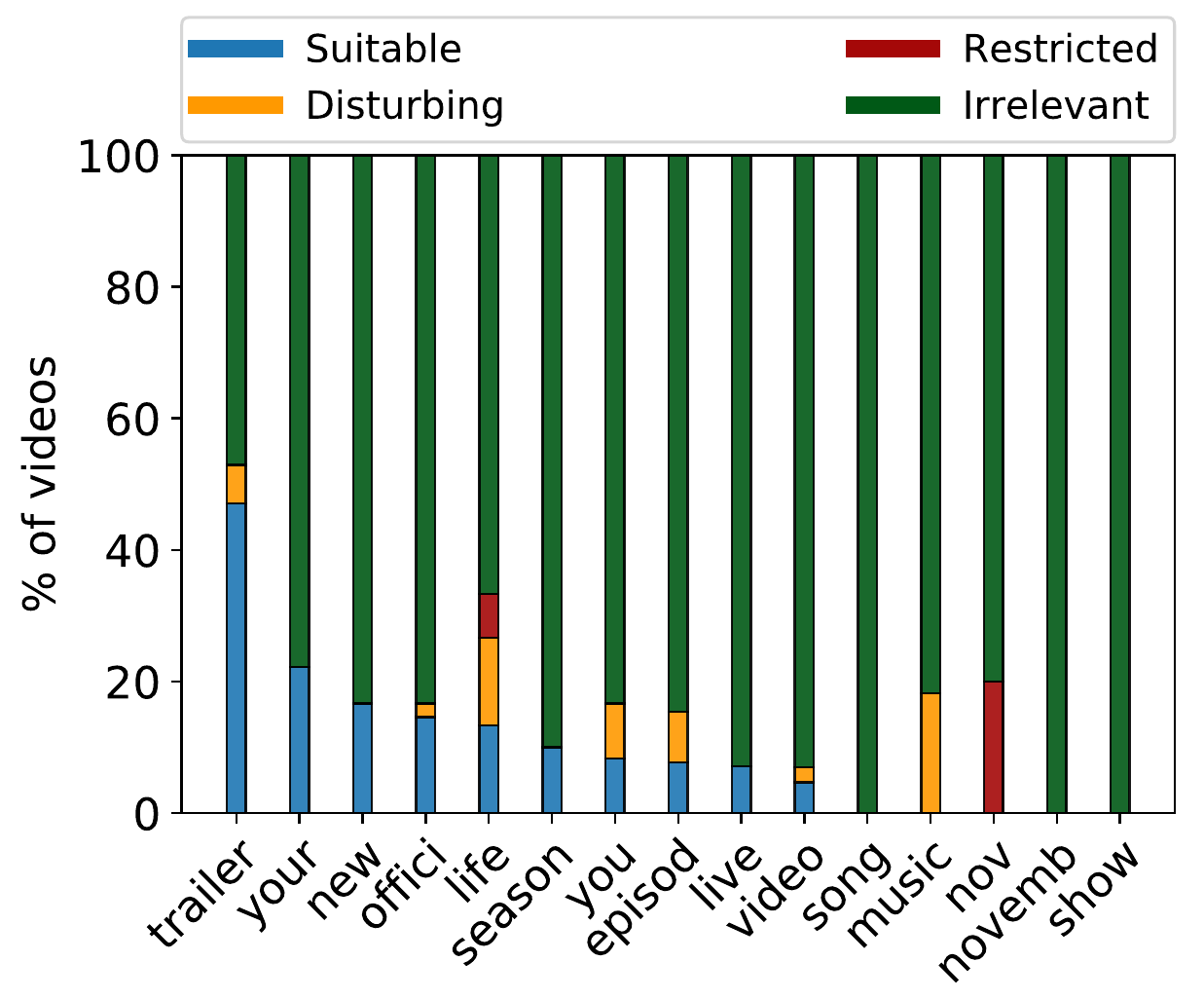}\label{fig:headline_normalized_mean_scores_popular}}
\caption{Per class proportion of videos for top 15 stems found in titles of (a) Elsagate-related; (b) other child-related; (c) random; and (d) popular videos.}
\label{fig:groundtruth_analysis_plots_headlines}
\end{figure*}

\begin{figure*}[t!]
\centering
\subfigure[]{\includegraphics[width=0.23\textwidth, height=1.4in]{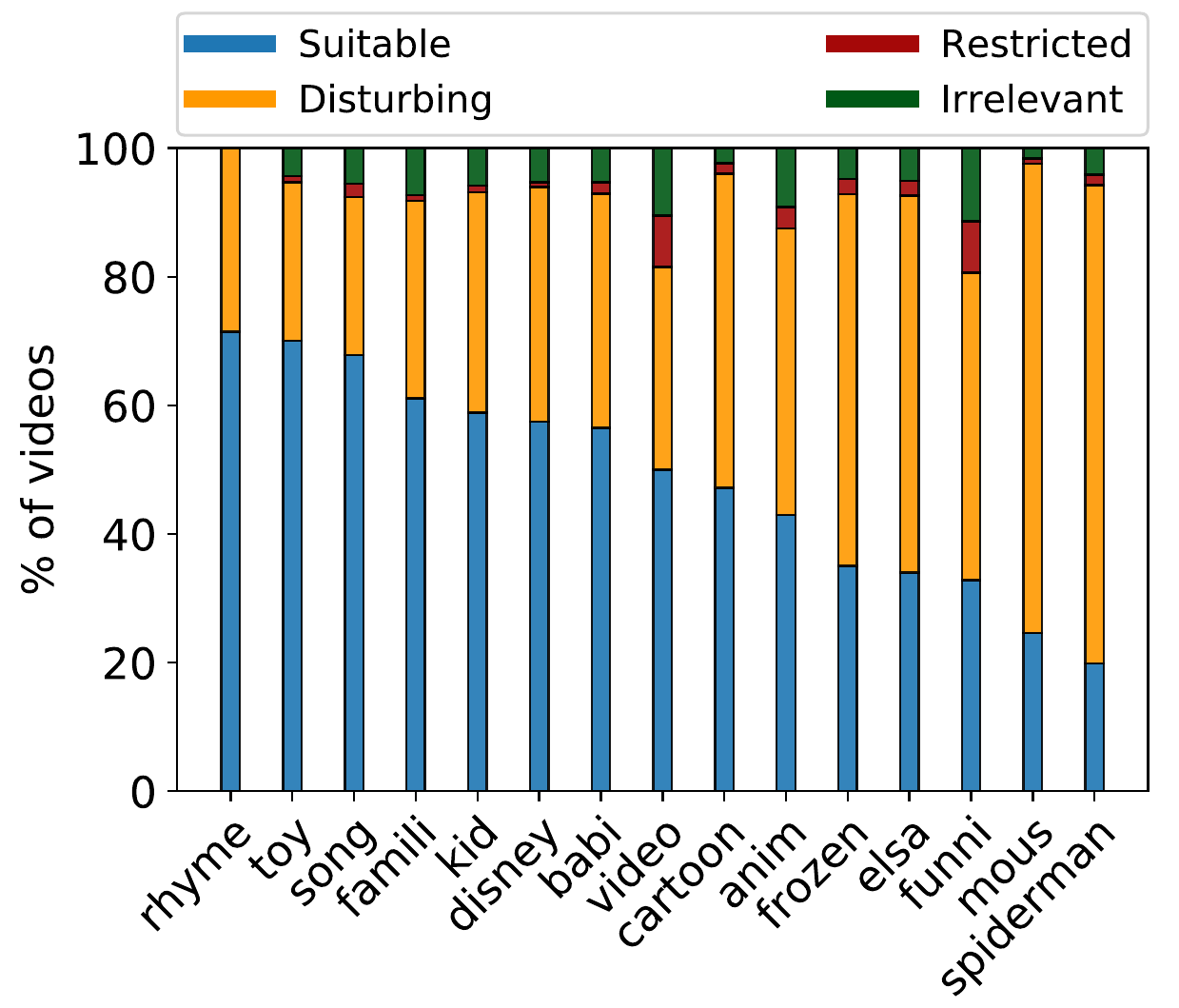}\label{fig:video_tags_stems_normalized_mean_scores_elsagate}}
\subfigure[]{\includegraphics[width=0.23\textwidth, height=1.4in]{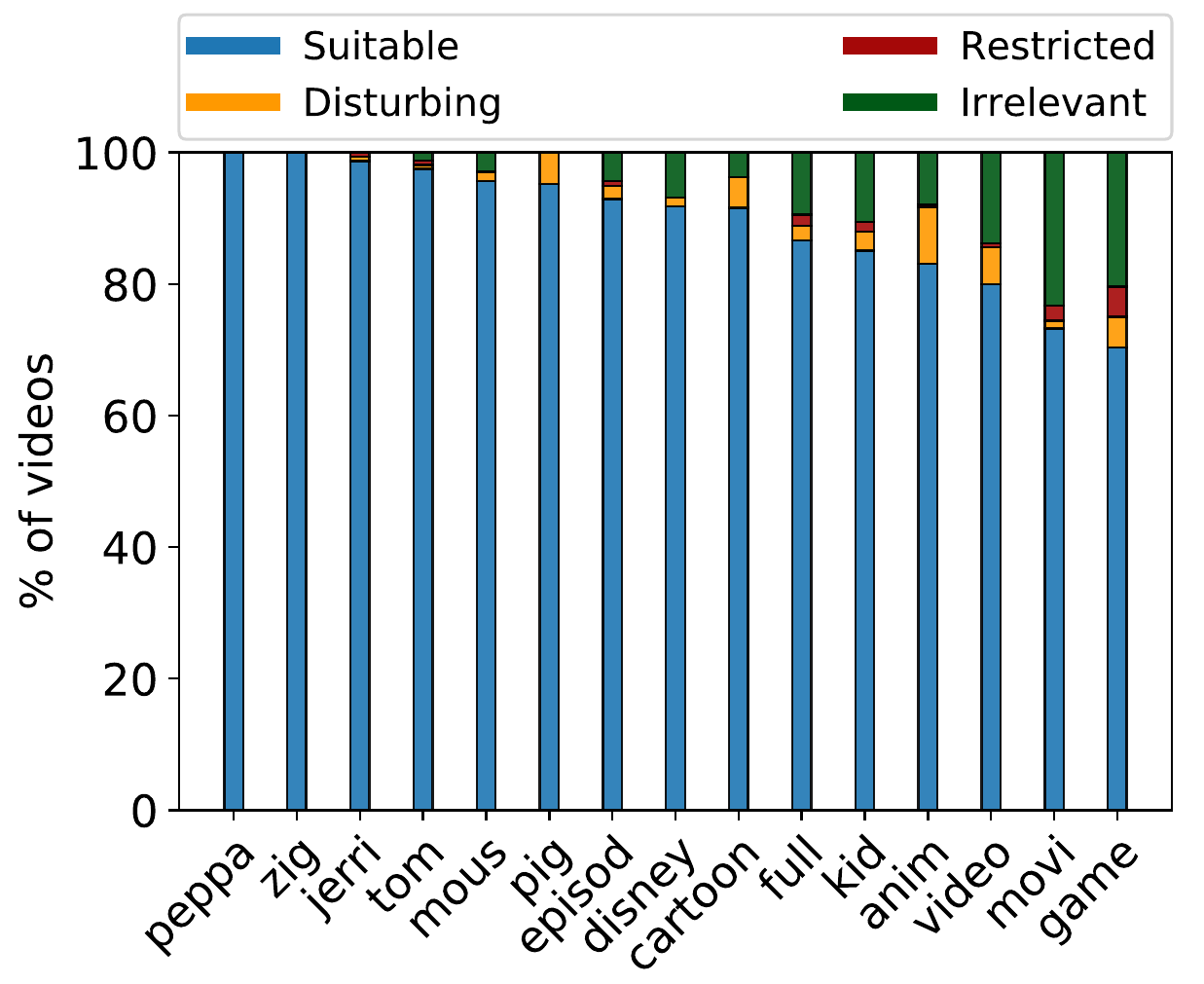}\label{fig:video_tags_stems_normalized_mean_scores_childrelated}}
\subfigure[]{\includegraphics[width=0.23\textwidth, height=1.4in]{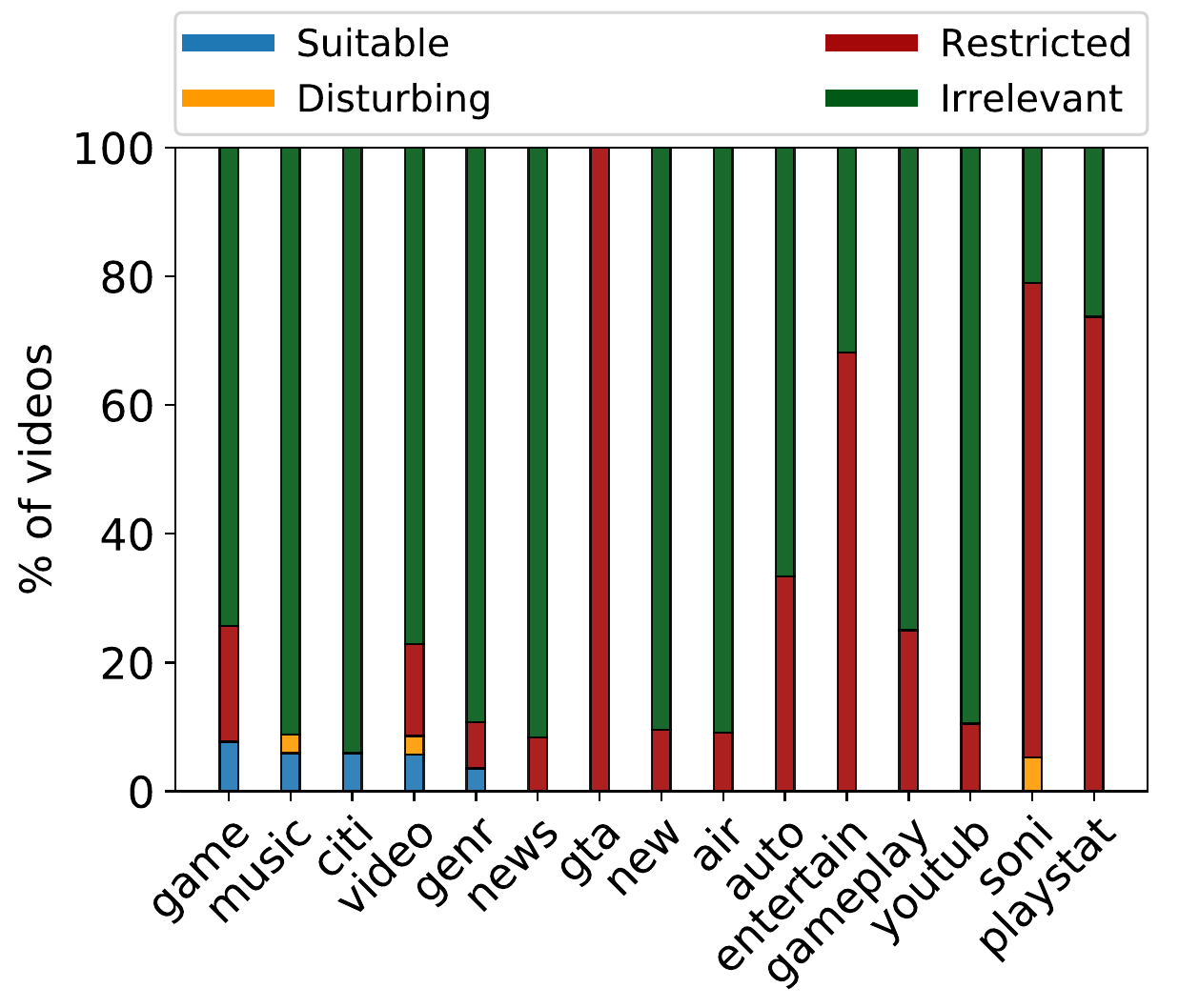}\label{fig:video_tags_stems_normalized_mean_scores_random}}
\subfigure[]{\includegraphics[width=0.23\textwidth, height=1.4in]{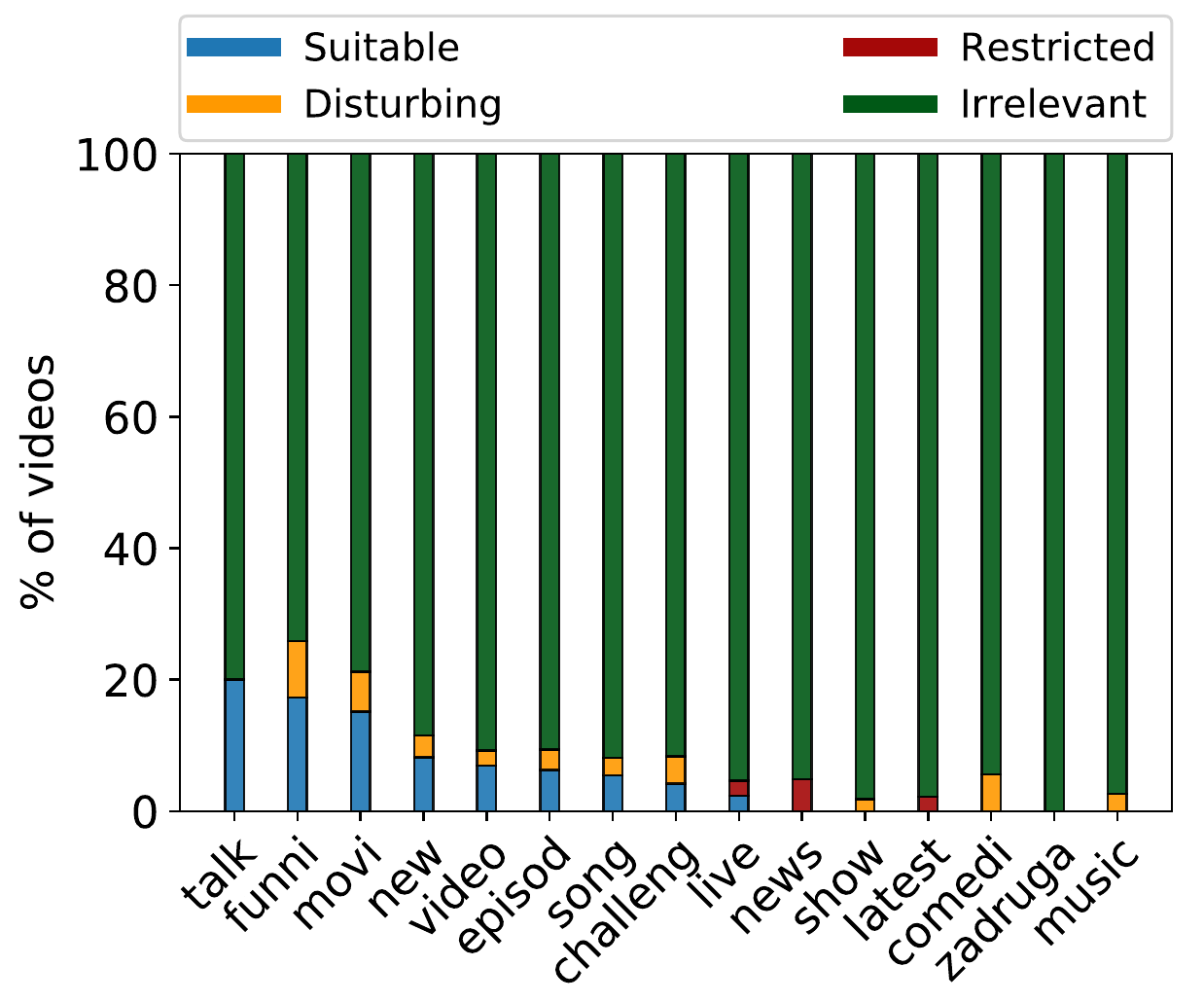}\label{fig:video_tags_stems_normalized_mean_scores_popular}}
\caption{Per class proportion of videos for the top 15 stems found in video tags of (a) Elsagate-related; (b) other child-related; (c) random; and (d) popular videos.}
\label{fig:groundtruth_analysis_plots_video_tags}
\end{figure*}

\begin{figure*}[t!]
\centering
\subfigure[]{\includegraphics[width=0.23\textwidth, height=1.45in]{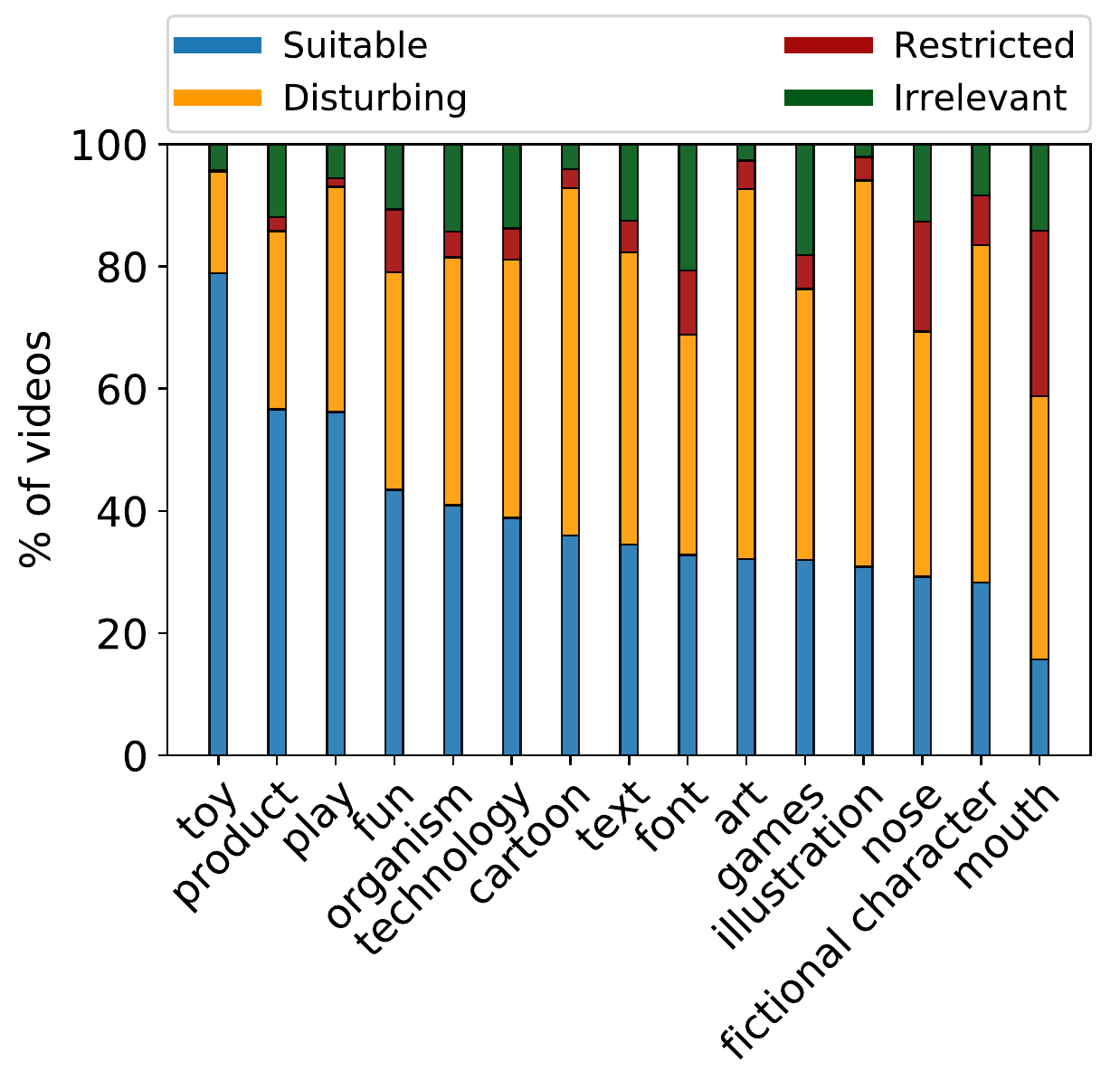}\label{fig:thumbnails_labels_normalized_mean_scores_elsagate}}
\subfigure[]{\includegraphics[width=0.23\textwidth, height=1.45in]{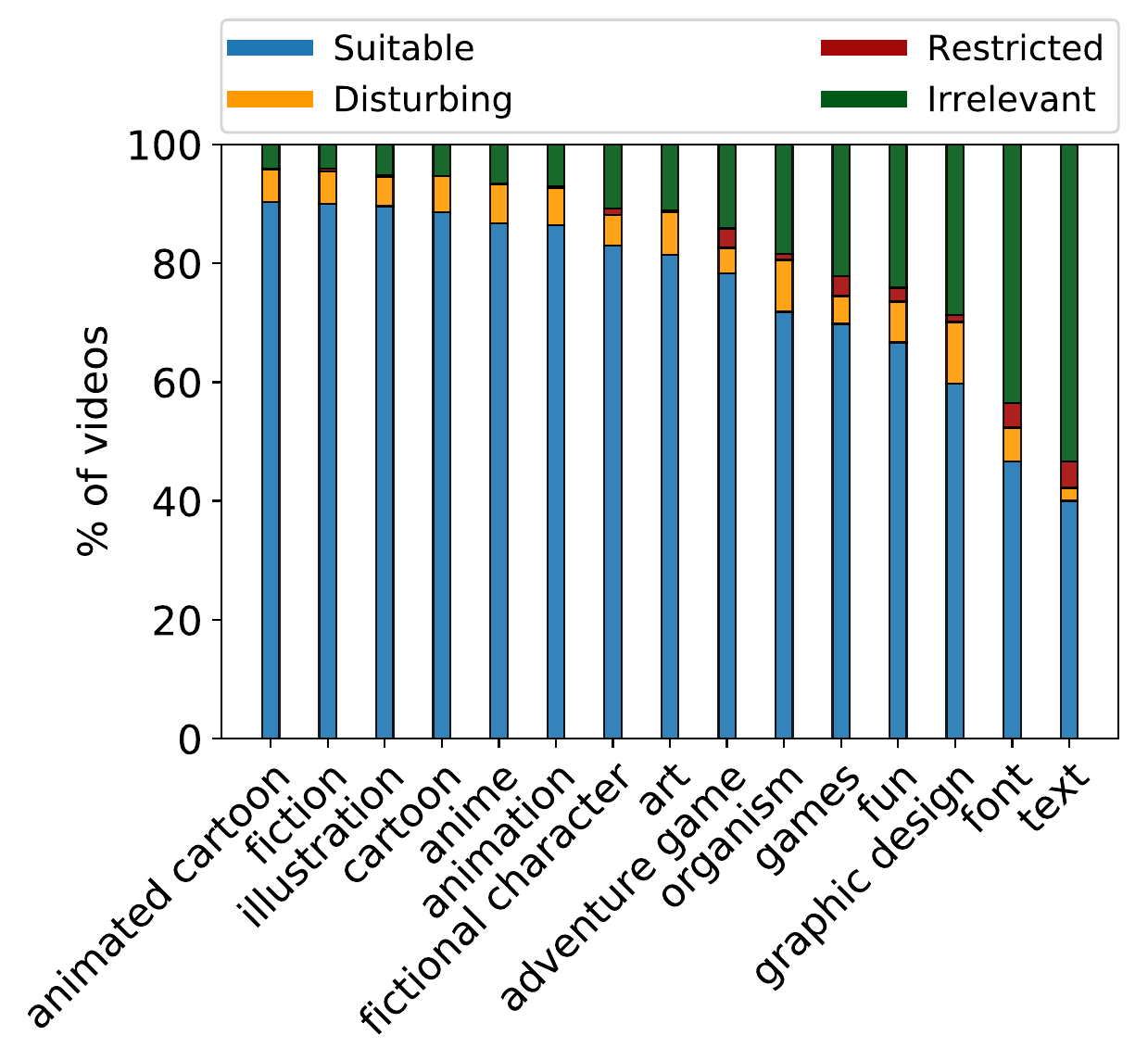}\label{fig:thumbnails_labels_normalized_mean_scores_childrelated}}
\subfigure[]{\includegraphics[width=0.23\textwidth, height=1.45in]{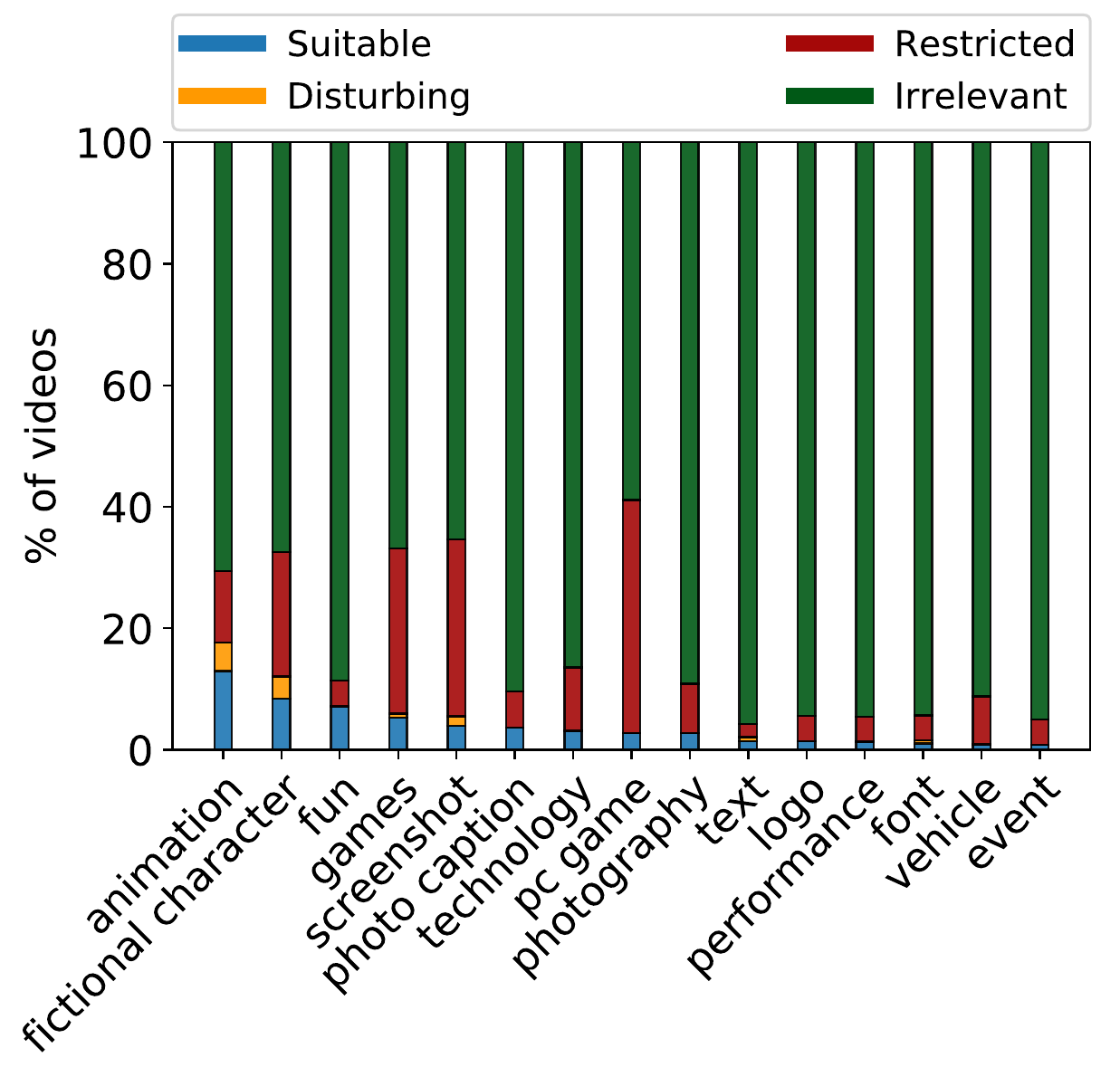}\label{fig:thumbnails_labels_normalized_mean_scores_random}}
\subfigure[]{\includegraphics[width=0.23\textwidth, height=1.45in]{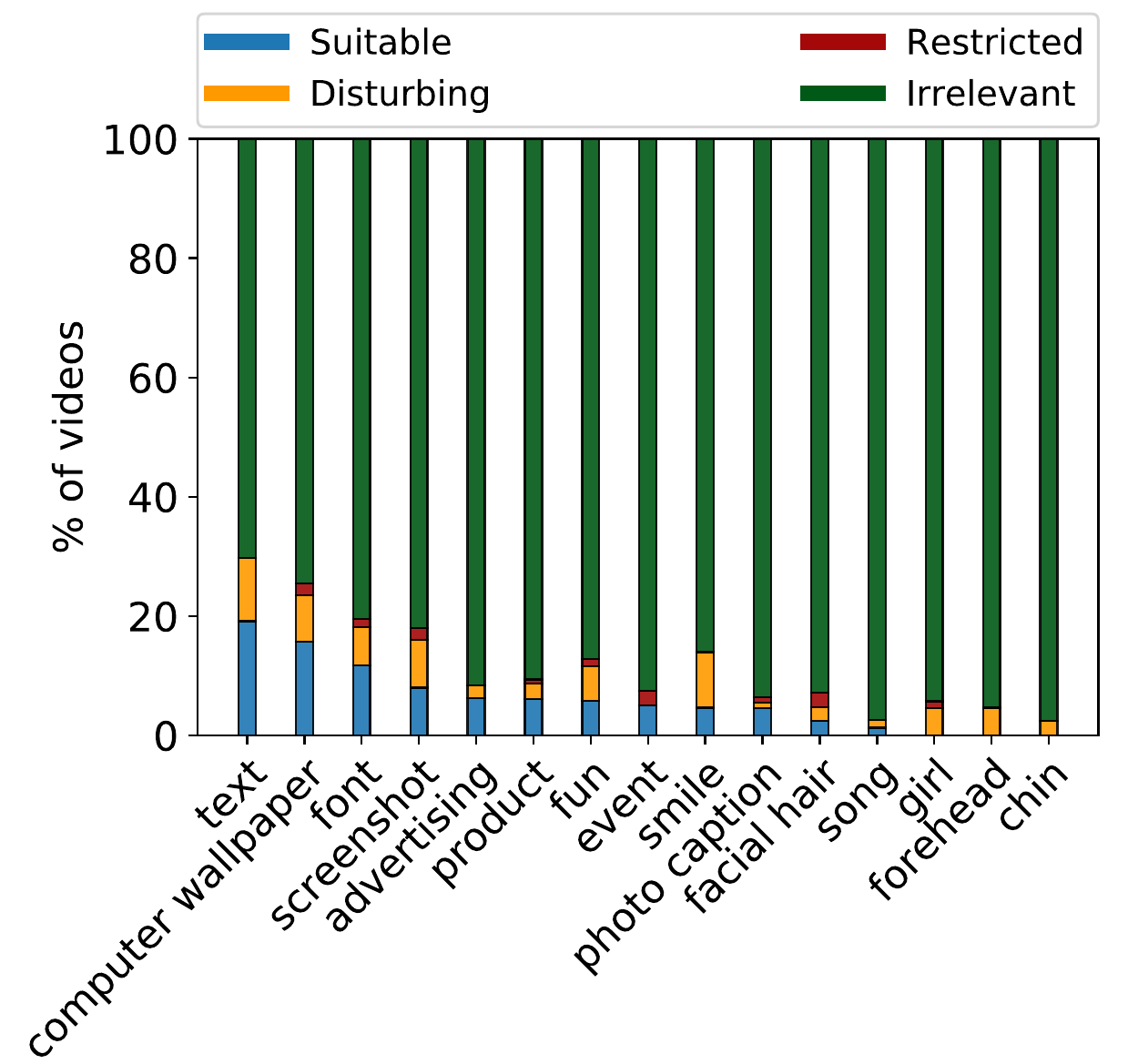}\label{fig:thumbnails_labels_normalized_mean_scores_popular}}
\caption{Per class proportion of videos for the top 15 labels found in thumbnails of (a) Elsagate-related; (b) other child-related; (c) random; and (d) popular videos.}
\label{fig:groundtruth_analysis_plots_thumbnails_stems}
\end{figure*}

\begin{figure*}[t!]
\centering
\subfigure[]{\includegraphics[width=0.23\textwidth, height=1.4in]{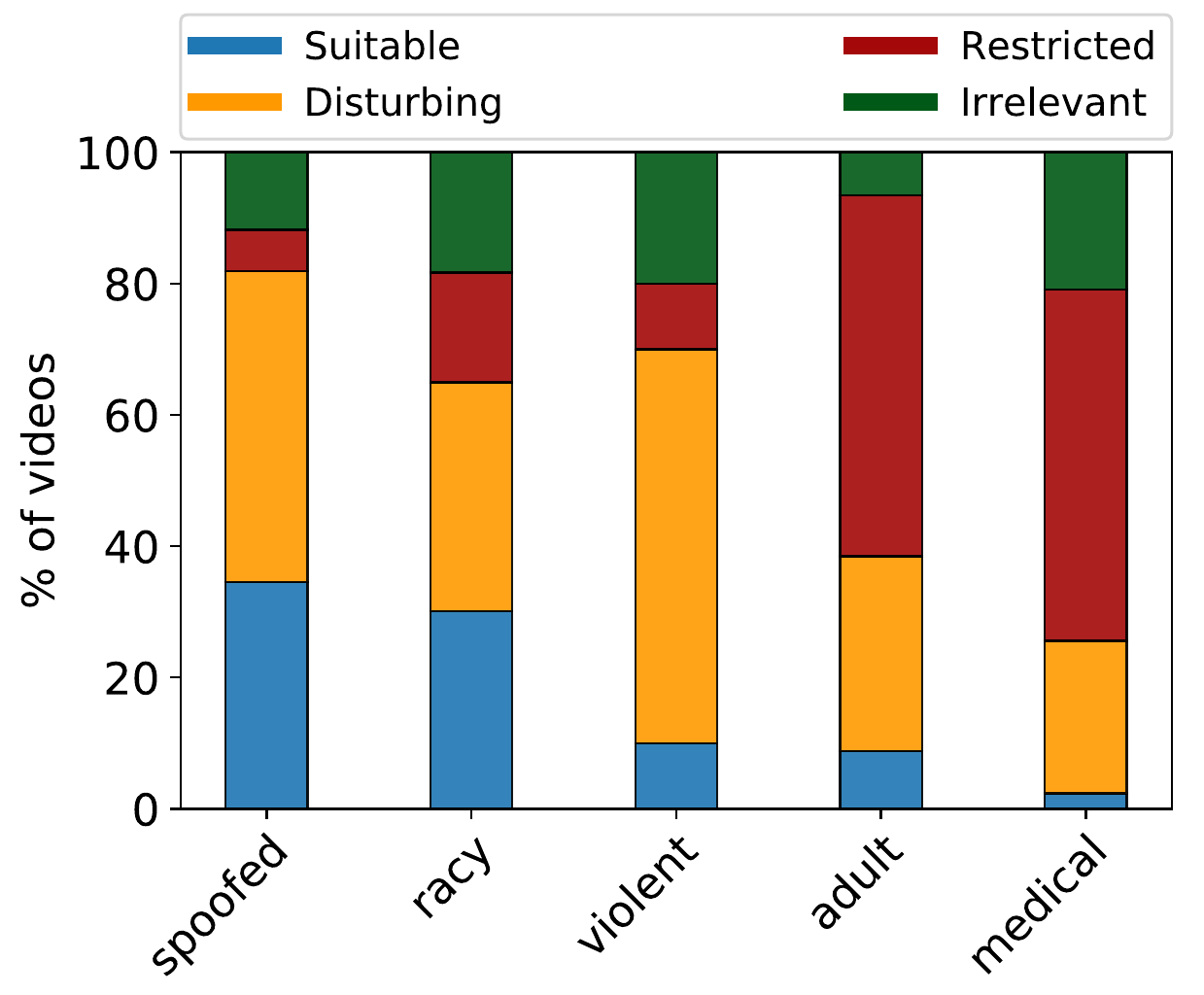}\label{fig:thumbnails_safe_search_normalized_mean_scores_elsagate}}
\subfigure[]{\includegraphics[width=0.23\textwidth, height=1.4in]{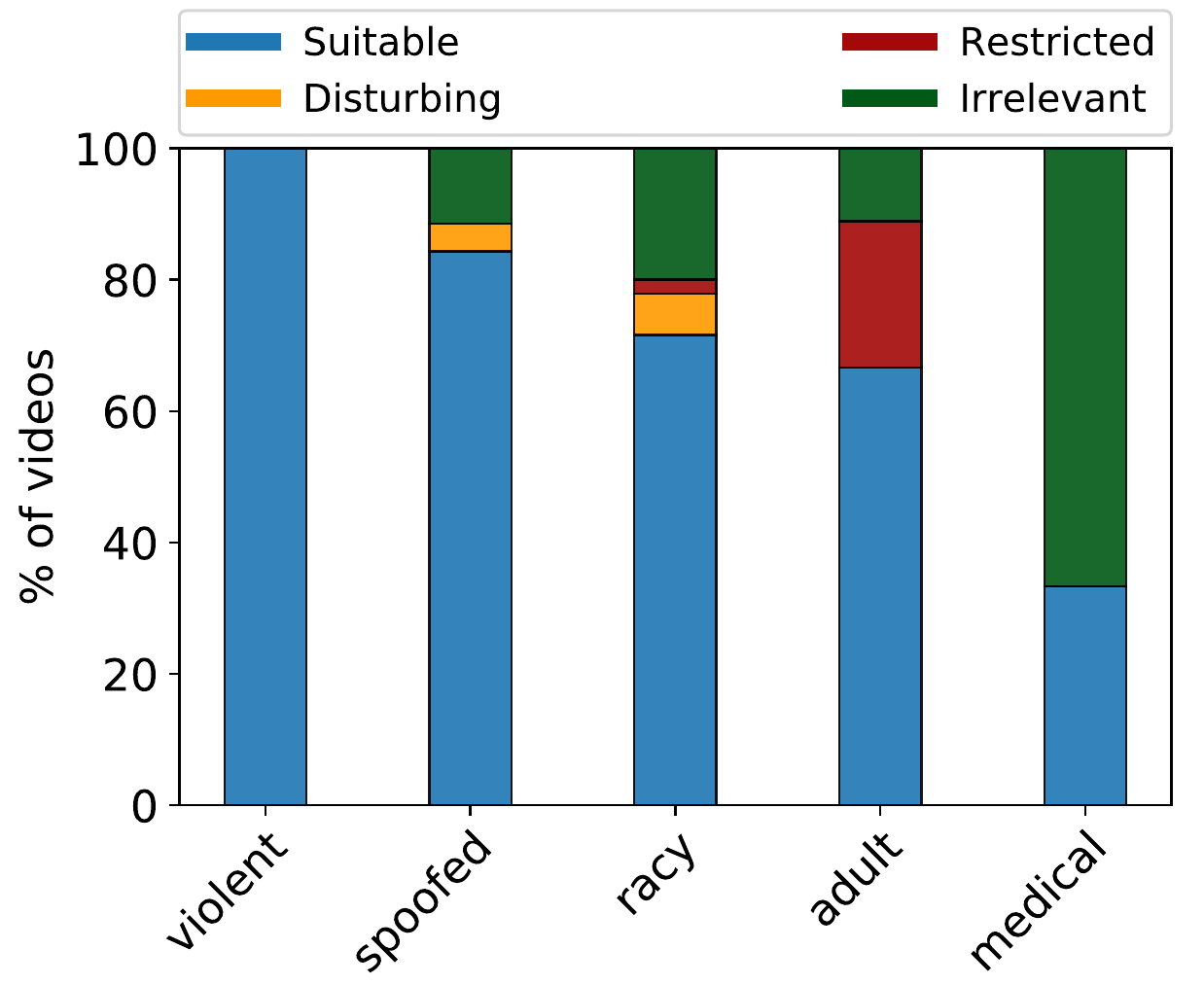}\label{fig:thumbnails_safe_search_normalized_mean_scores_childrelated}}
\subfigure[]{\includegraphics[width=0.23\textwidth, height=1.4in]{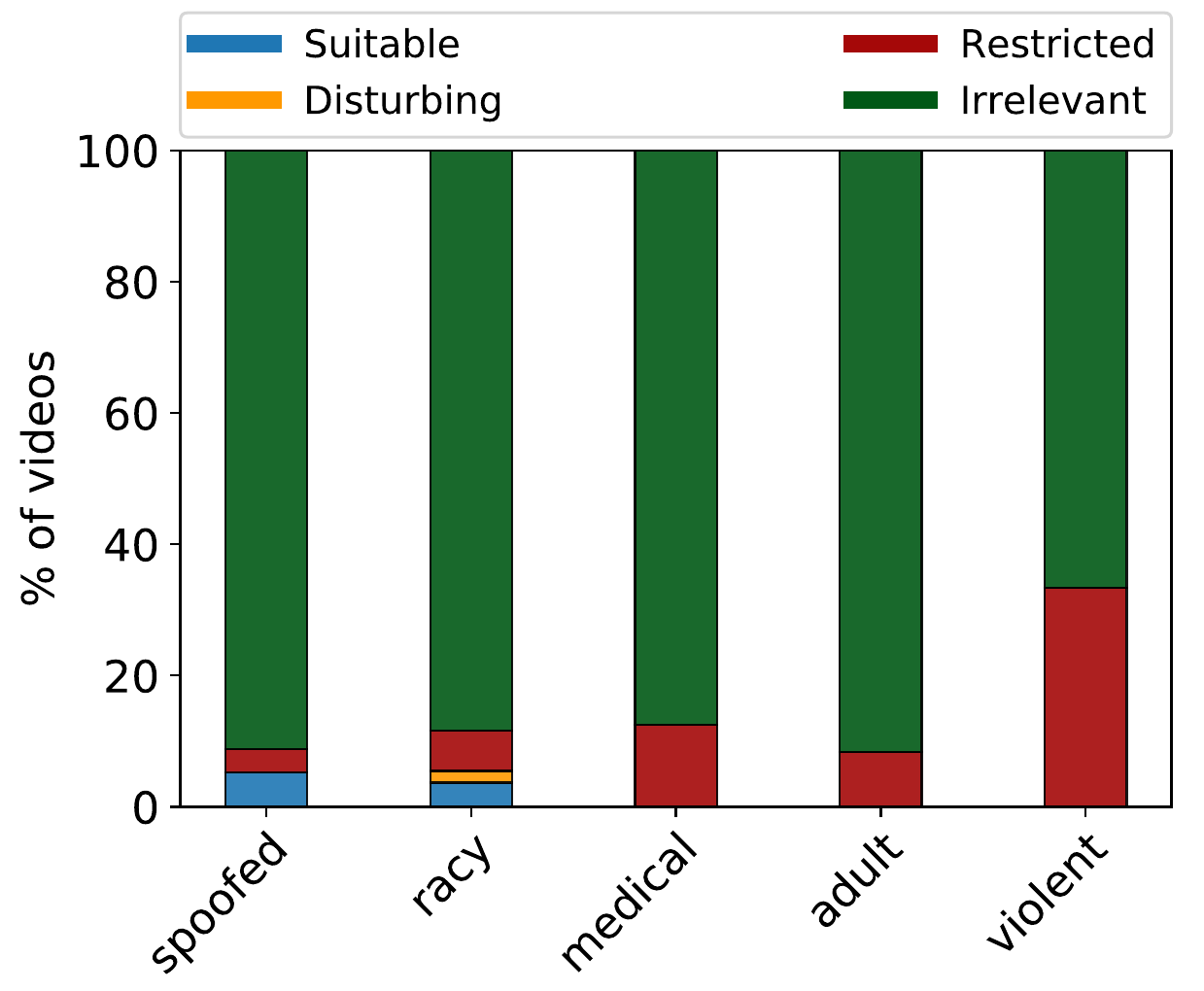}\label{fig:thumbnails_safe_search_normalized_mean_scores_random}}
\subfigure[]{\includegraphics[width=0.23\textwidth, height=1.4in]{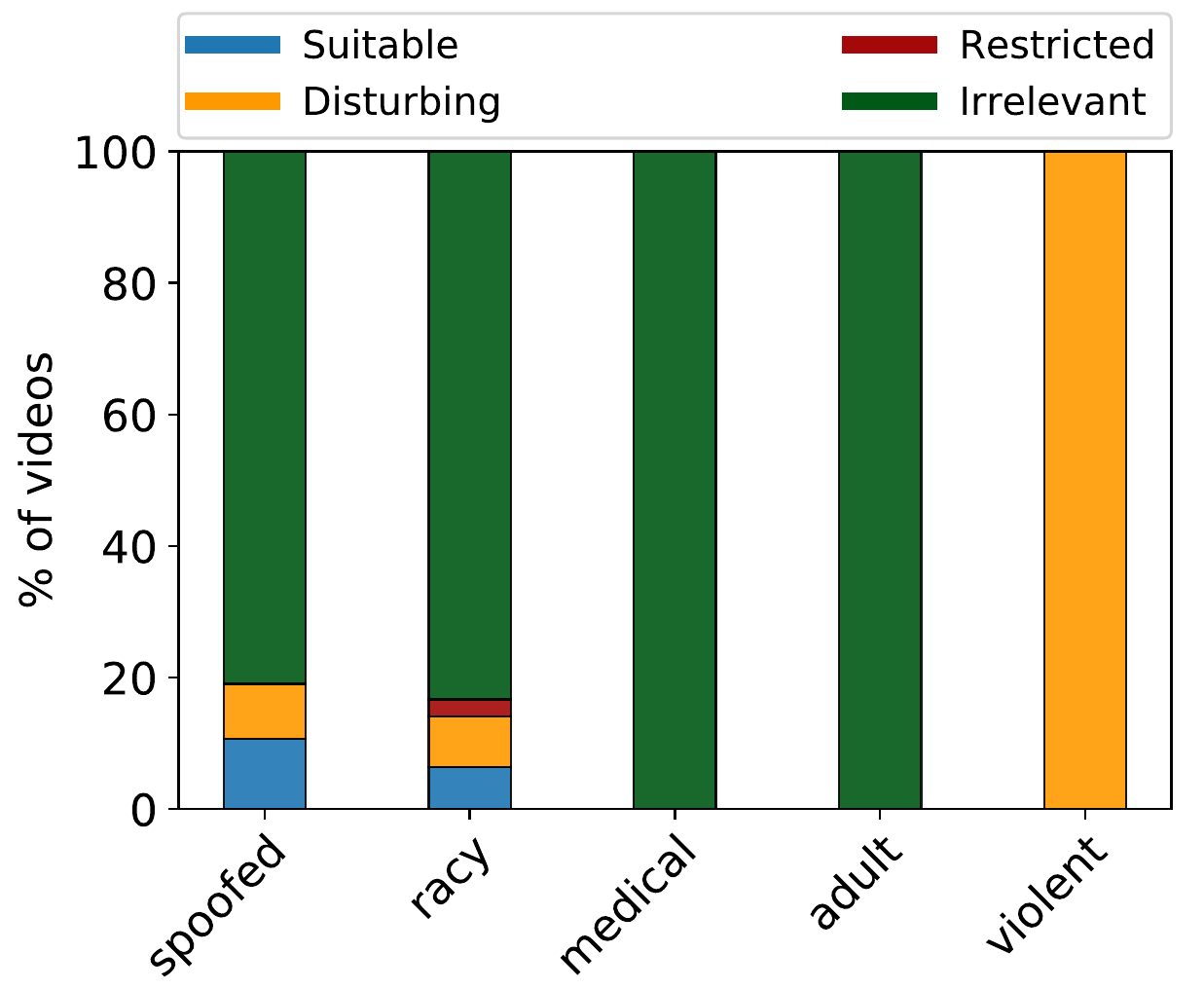}\label{fig:thumbnails_safe_search_normalized_mean_scores_popular}}
\caption{Per class proportion of videos that their thumbnail contains spoofed, adult, medical, violent, and/or racy content for (a) Elsagate-related; (b) other child-related; (c) random; and (d) popular videos.}
\label{fig:groundtruth_analysis_plots_thumbnails_safesearch_categories}
\end{figure*}

\begin{figure*}[t!]
\centering
\subfigure[]{\includegraphics[width=0.23\textwidth, height=1.2in]{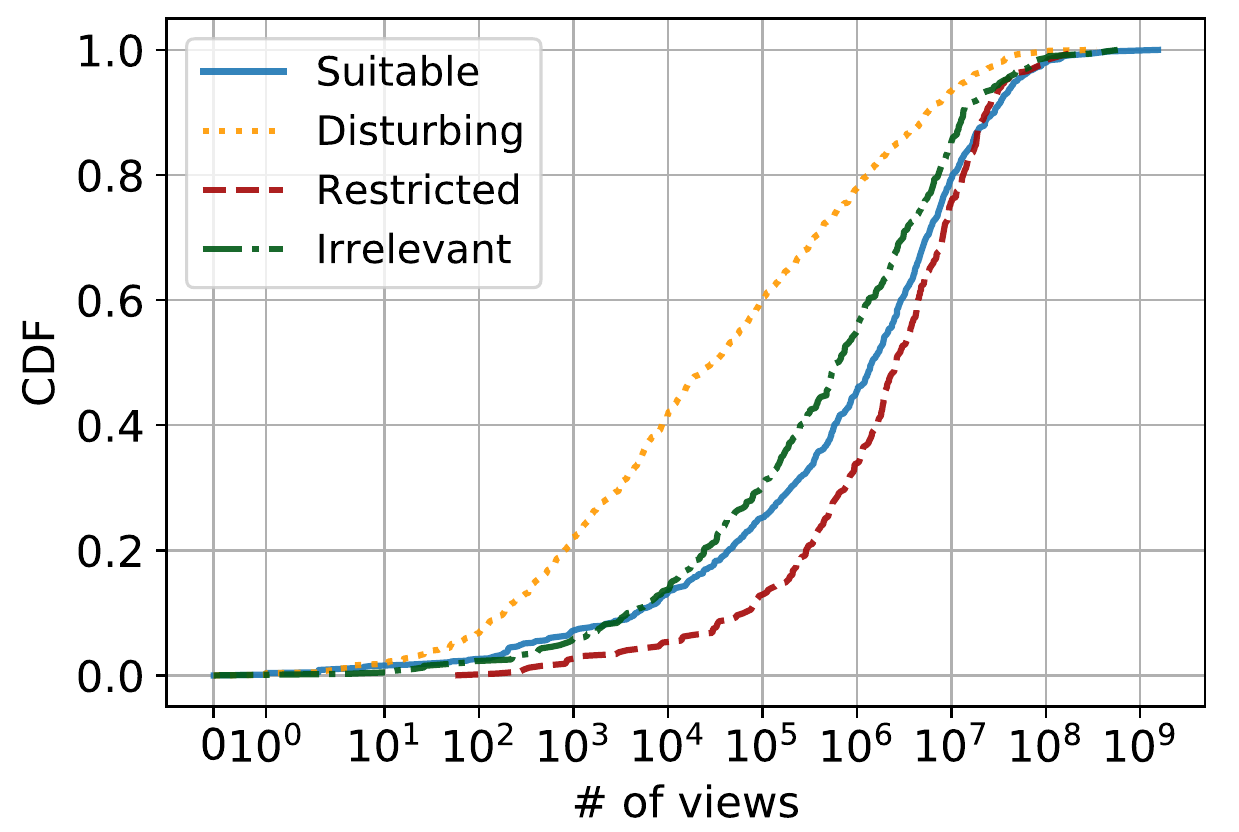}}
\subfigure[]{\includegraphics[width=0.23\textwidth, height=1.2in]{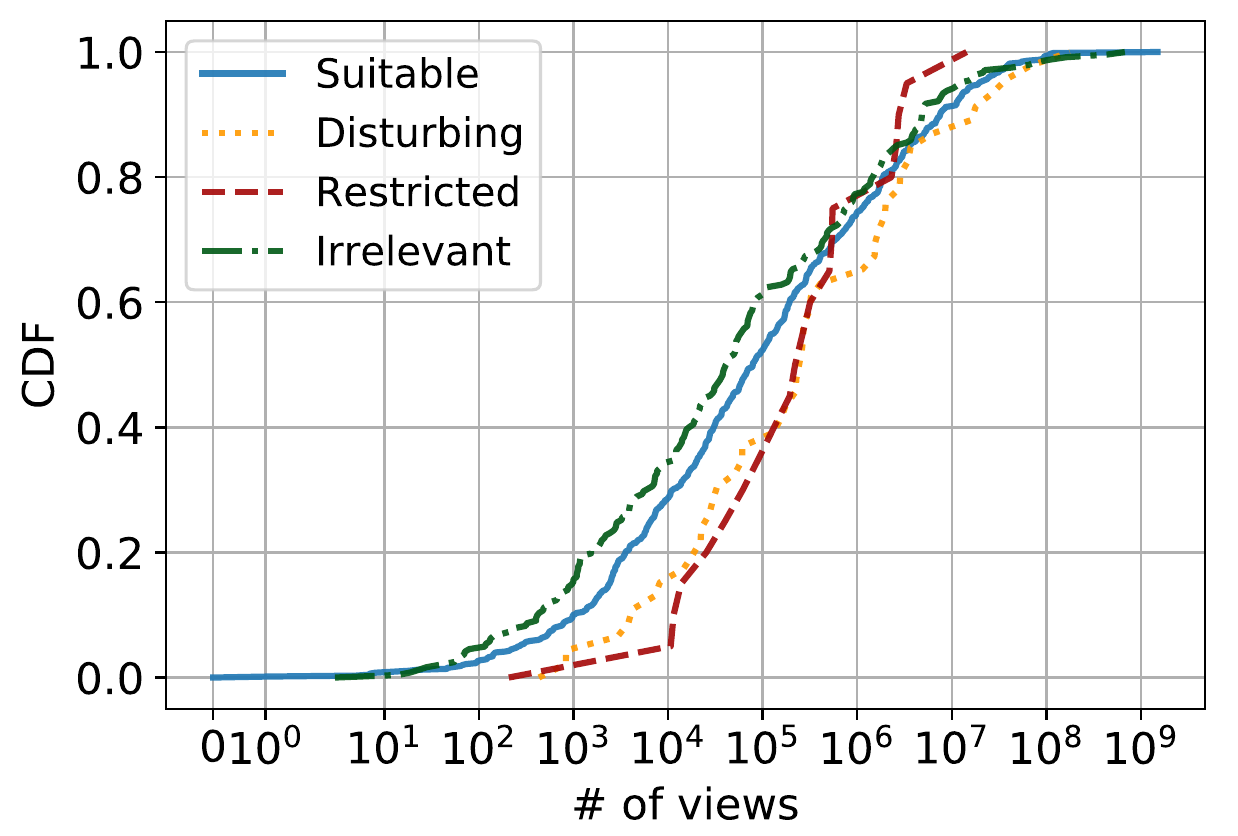}}
\subfigure[]{\includegraphics[width=0.23\textwidth, height=1.2in]{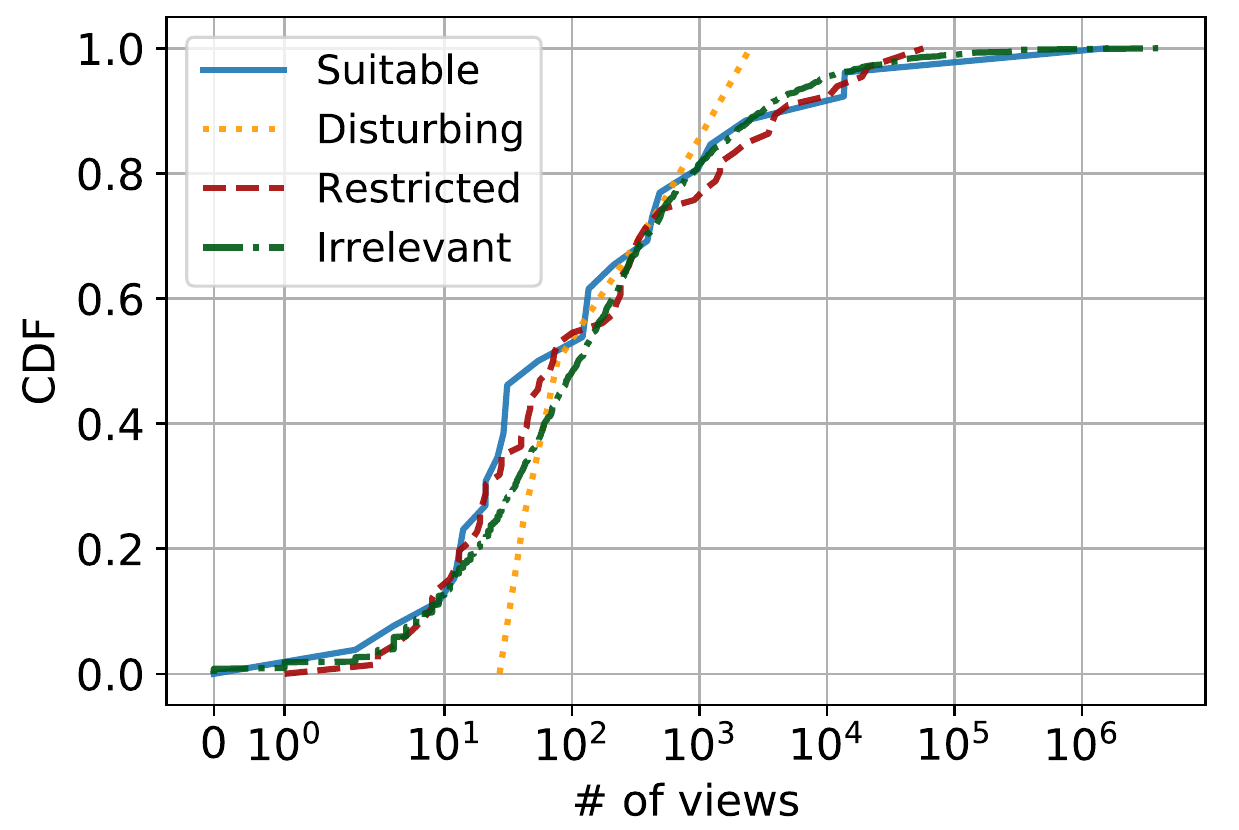}}
\subfigure[]{\includegraphics[width=0.23\textwidth, height=1.2in]{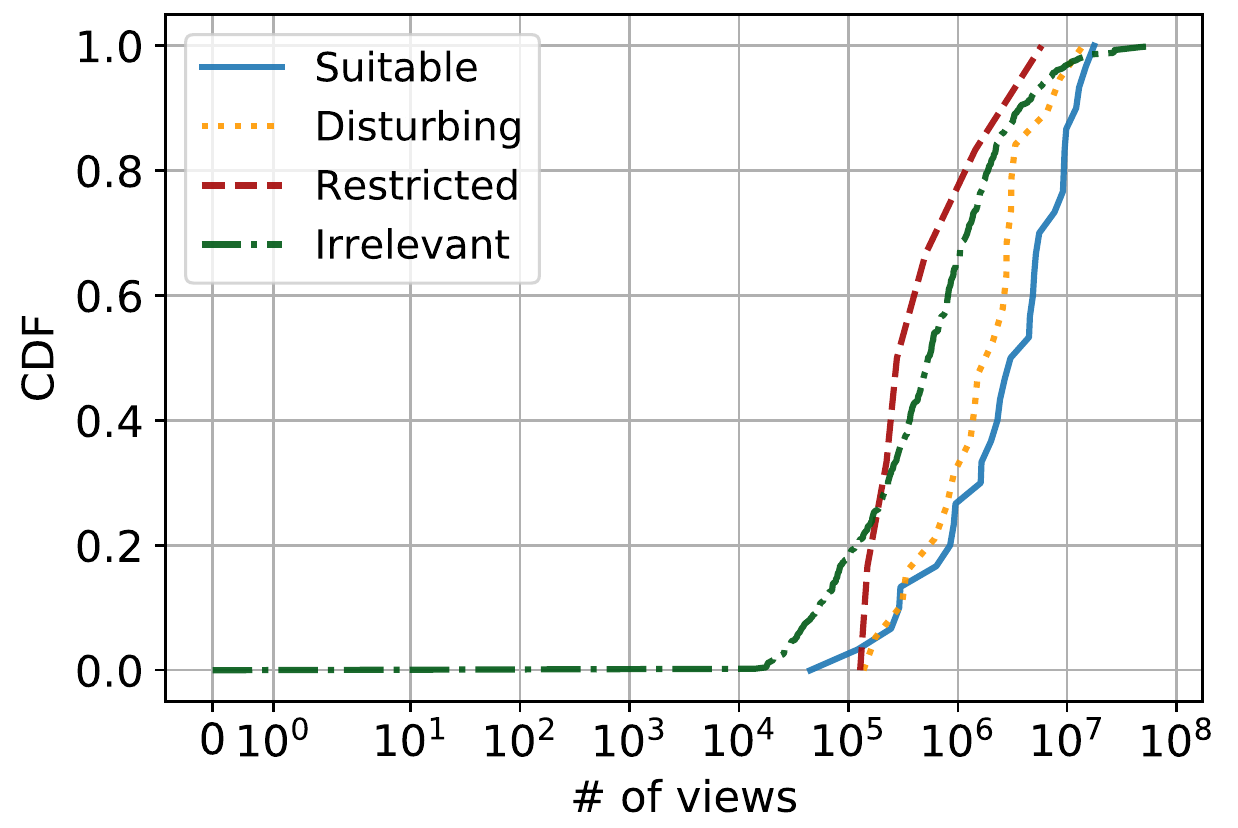}}
\caption{CDF of the number of views per class for (a) Elsagate-related (b) other child-related, (c) random, and (d) popular videos.}
\label{fig:cdf_plots_views}
\end{figure*}

\begin{figure*}[t!]
\centering
\subfigure[]{\includegraphics[width=0.23\textwidth, height=1.2in]{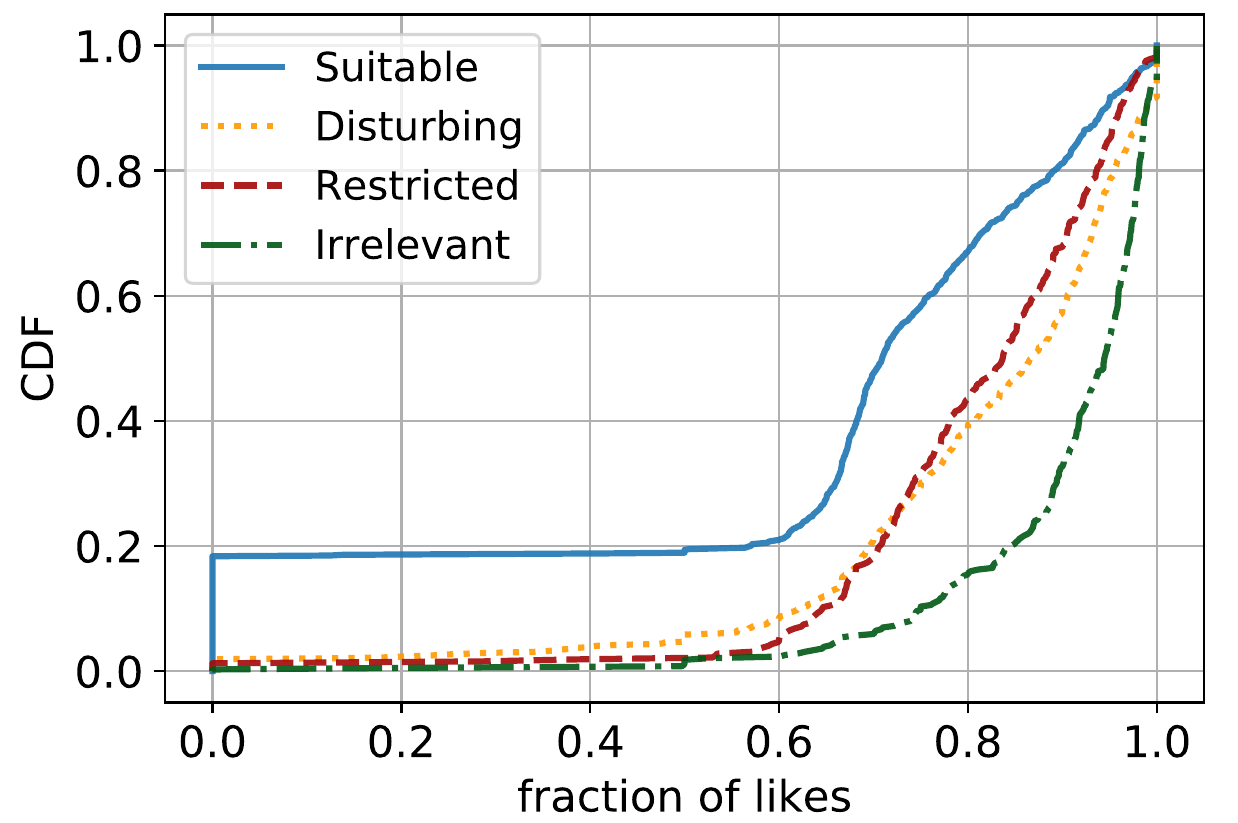}}
\subfigure[]{\includegraphics[width=0.23\textwidth, height=1.2in]{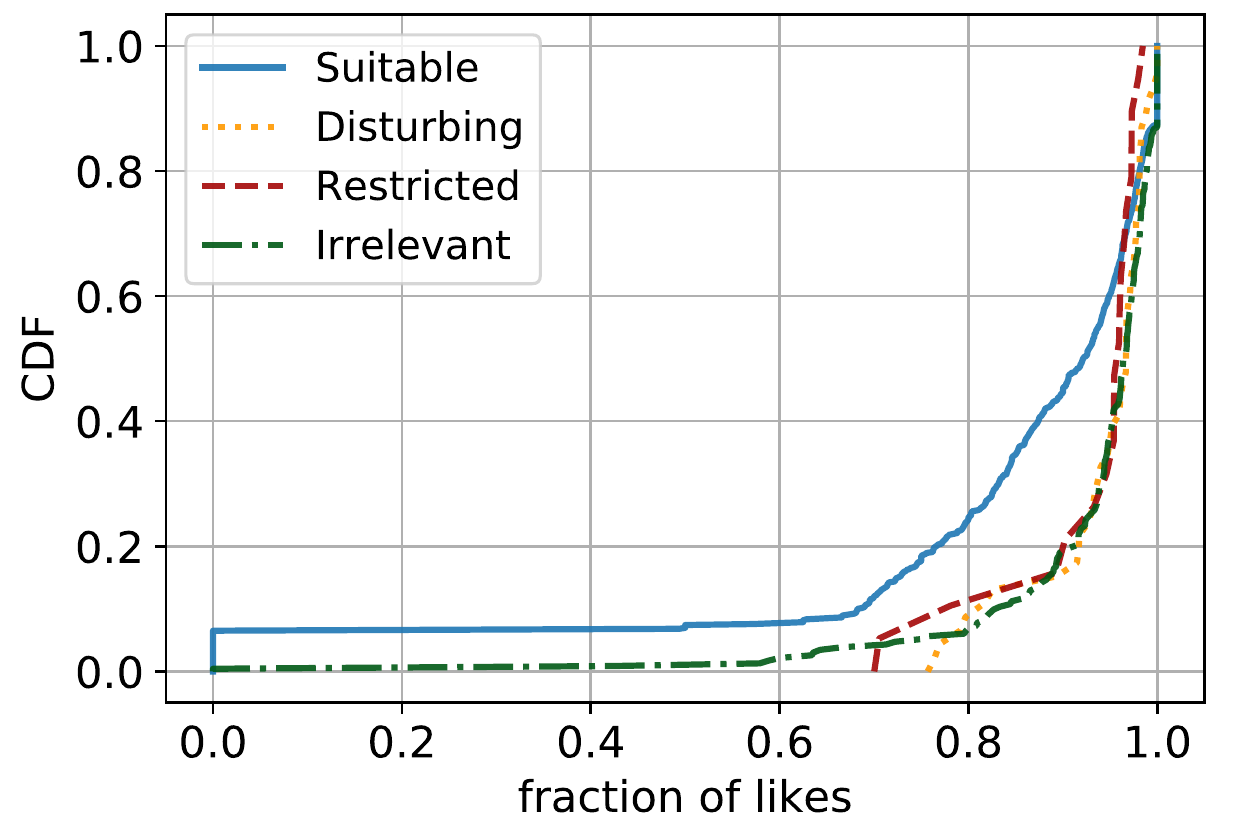}}
\subfigure[]{\includegraphics[width=0.23\textwidth, height=1.2in]{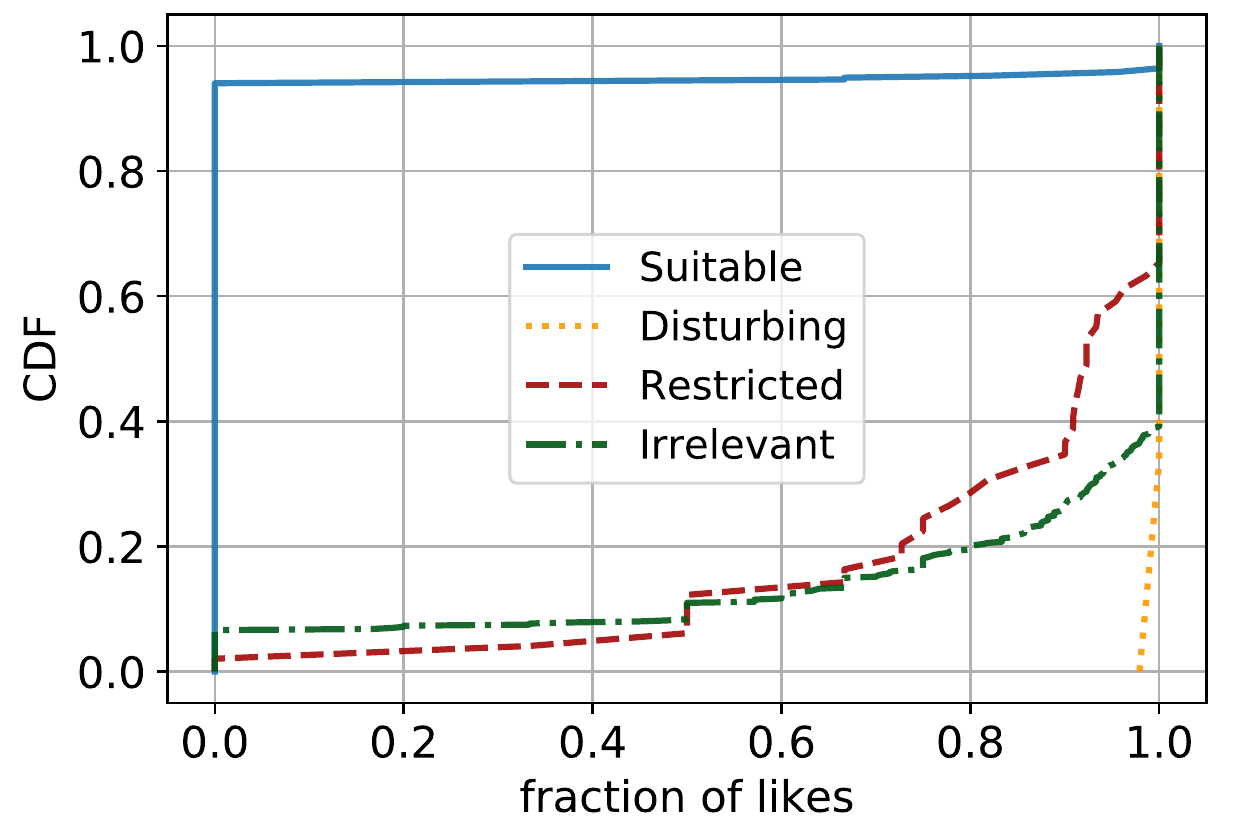}}
\subfigure[]{\includegraphics[width=0.23\textwidth, height=1.2in]{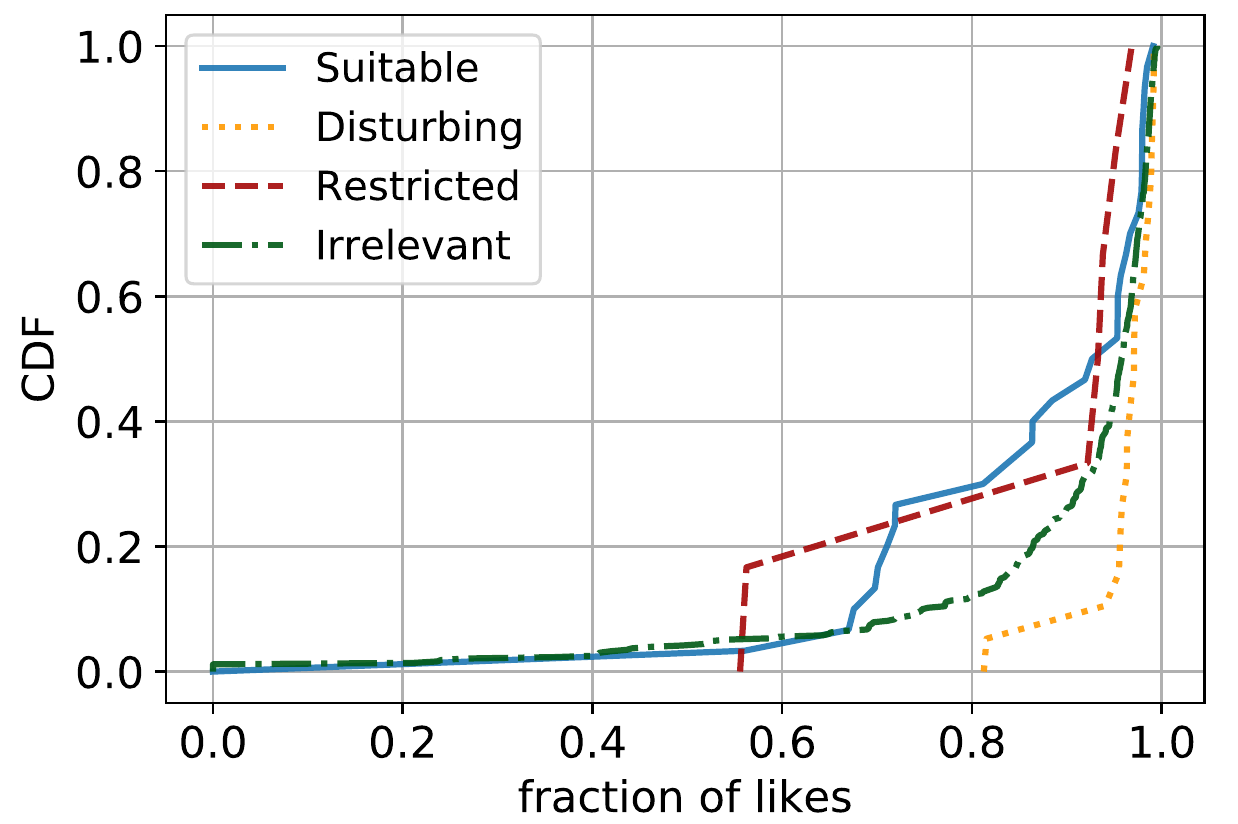}}
\caption{CDF of the fraction of likes to dislikes per class for (a) Elsagate-related (b) other child-related, (c) random, and (d) popular videos.}
\label{fig:cdf_plots_likesdislikes_fraction}
\end{figure*}

\begin{figure*}[t!]
\centering
\subfigure[]{\includegraphics[width=0.23\textwidth, height=1.2in]{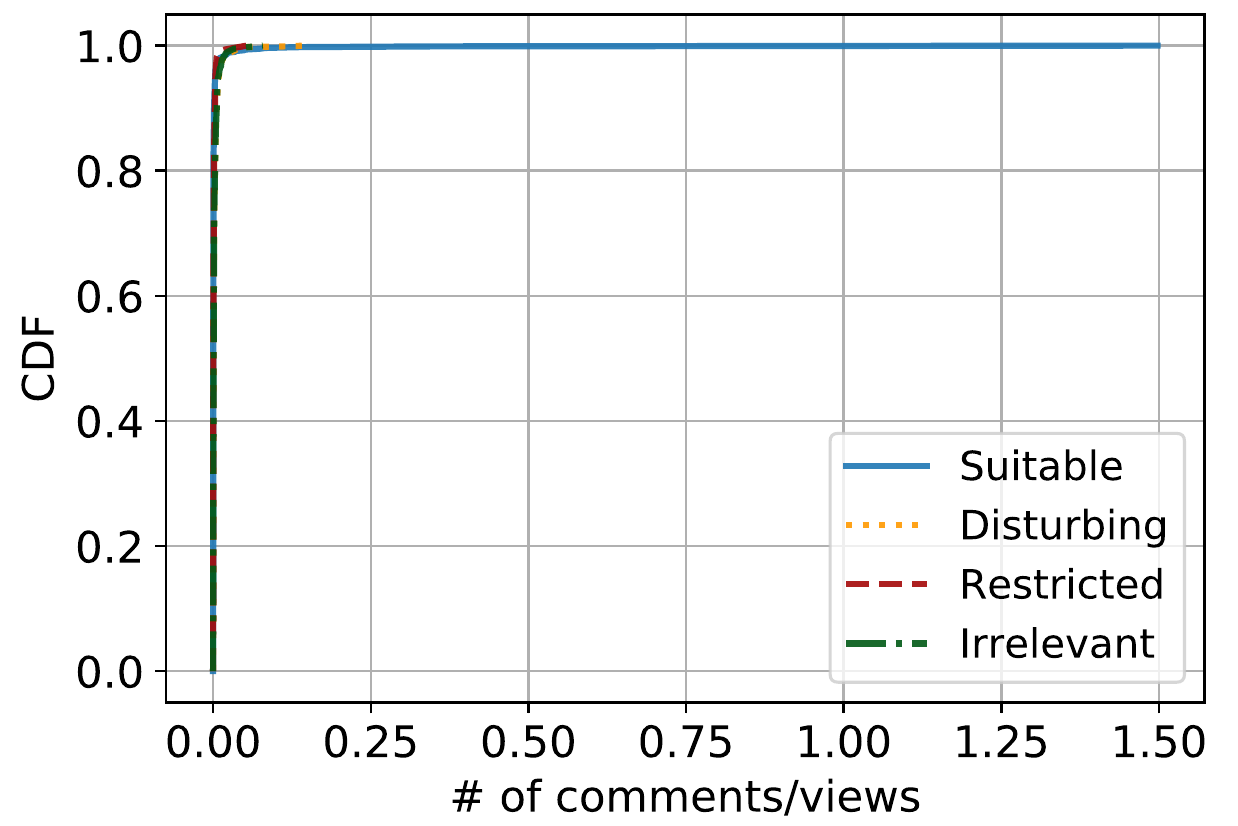}}
\subfigure[]{\includegraphics[width=0.23\textwidth, height=1.2in]{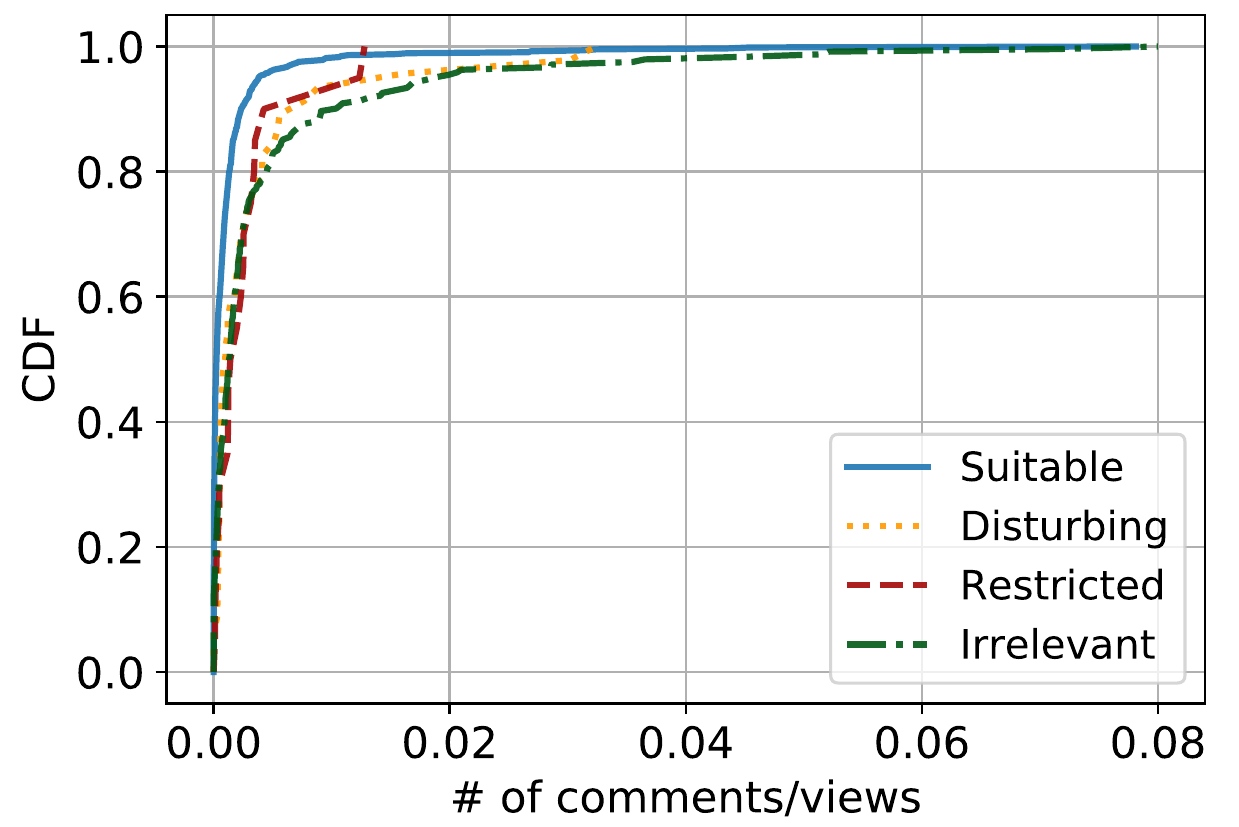}}
\subfigure[]{\includegraphics[width=0.23\textwidth, height=1.2in]{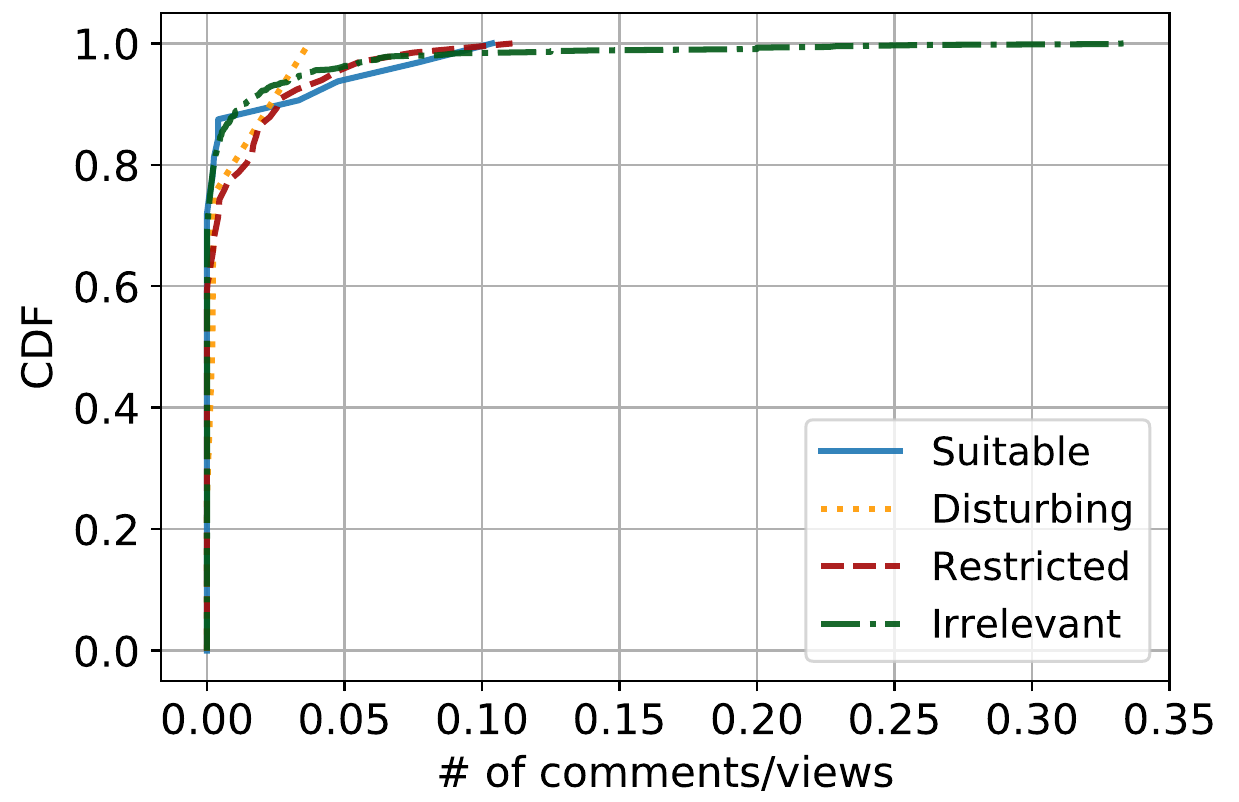}}
\subfigure[]{\includegraphics[width=0.23\textwidth, height=1.2in]{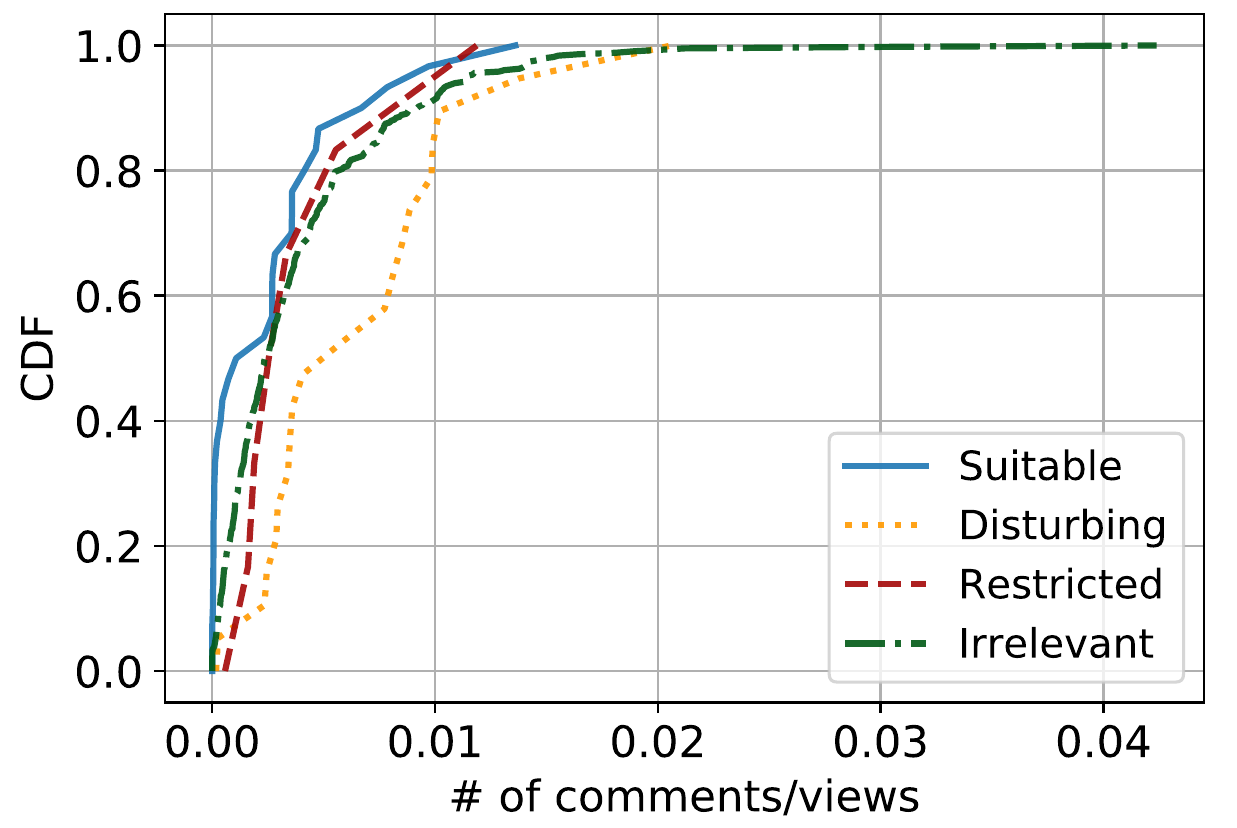}}
\caption{CDF of the number of comments/views per class for (a) Elsagate-related (b) other child-related, (c) random, and (d) popular videos.}
\label{fig:cdf_plots_commentsviews_fraction}
\end{figure*}

\begin{table}[t!]
\centering
\resizebox{\columnwidth}{!}{%
\begin{tabular}{llrrrr}
\toprule
\textbf{} & \textbf{Category} & \multicolumn{1}{c}{\textbf{Suitable}} & \multicolumn{1}{l}{\textbf{Disturbing}} & \multicolumn{1}{l}{\textbf{Restricted}} & \multicolumn{1}{l}{\textbf{Irrelevant}} \\ \midrule
Elsagate- & EN & 353 (44\%) & 208 (24\%) & 99 (31\%) & 98 (25\%) \\
related & F\&A & 130 (16\%) & 190 (22\%) & 39 (12\%) & 33 (8\%) \\
 & ED & 128 (16\%) & 21 (3\%) & 17 (5\%) & 16 (4\%) \\
 & P\&B & 109 (13\%) & 239 (28\%) & 71 (22\%) & 73 (19\%) \\
 & M & 21 (3\%) & 15 (2\%) & 8 (3\%) & 45 (11\%) \\ \midrule

Other & EN & 131 (20\%) & 7 (15\%) & 9 (43\%) & 51 (21\%) \\
Child- & F\&A & 317 (49\%) & 27 (58\%) & 3 (14\%) & 26 (11\%) \\
related & ED & 27 (4\%) & 1 (2\%) & 2 (10\%) & 34 (14\%) \\
 & P\&B & 130 (20\%) & 4 (8\%) & 2 (9\%) & 35 (14\%) \\
 & M & 5 (1\%) & 2 (4\%) & 0 (0\%) & 26 (11\%) \\ \midrule

Random & EN & 4 (15\%) & 1 (20\%) & 3 (5\%) & 68 (8\%) \\
 & F\&A & 1 (4\%) & 1 (20\%) & 1 (2\%) & 18 (2\%) \\
 & ED & 1 (4\%) & 0 (0\%) & 0 (0\%) & 31 (4\%) \\
 & P\&B & 13 (48\%) & 3 (60\%) & 21 (31\%) & 354 (41\%) \\
 & M & 1 (3\%) & 0 (0\%) & 0 (0\%) & 79 (9\%) \\ \midrule

Popular & EN & 12 (39\%) & 9 (45\%) & 2 (29\%) & 168 (39\%) \\
 & F\&A & 9 (29\%) & 7 (35\%) & 0 (0\%) & 13 (3\%) \\
 & ED & 2 (7\%) & 0 (0\%) & 0 (0\%) & 11 (3\%) \\
 & P\&B & 2 (6\%) & 0 (0\%) & 0 (0\%) & 32 (7\%) \\
 & M & 0 (0\%) & 1 (5\%) & 0 (0\%) & 63 (15\%) \\ \midrule
\end{tabular}%
}
\caption{Number of videos in each category per class for each subset of videos in our ground truth dataset. EN: Entertainment, F\&A: Film \& Animation, ED: Education, P\&B: People \& Blogs, M: Music.}
\label{tab:groundtruth_video_categories_stats}
\end{table}

\subsection{Ground Truth Dataset Analysis}

\noindent \textbf{Category.}
First, we look at the categories of the videos in our ground truth dataset. 
Table~\ref{tab:groundtruth_video_categories_stats} reports the top five categories, for each subset of videos. 
Most of the disturbing and restricted in the Elsagate-related videos are in Entertainment (24\% and 31\%), People \& Blogs (28\% and 22\%), and Film \& Animation (22\% and 12\%). 
These results are similar with previous work~\cite{chaudhary2013contextual}. 
A similar trend is also observed in all the other sets of videos.
In addition, in the Elsagate-related videos we find a non-negligible percentage of disturbing videos in seemingly innocent categories like Education (3\%) and Music (2\%). 
This is alarming since it indicates that disturbing videos ``infiltrate'' categories of videos that are likely to be selected by the toddler's parents. 
Unsurprisingly, after manually inspecting all the disturbing videos in the Education and Music categories, we find that the majority of them are nursery rhymes, ``wrong heads'', and ``peppa pig'' videos with disturbing content.

\noindent \textbf{Title.} 
The title of a video is an important factor that affects whether a particular video will be recommended when viewing other toddler-oriented videos.
Consequently, we study the titles in our ground truth dataset to understand the tactics and terms that are usually used by uploaders of disturbing or 
restricted videos on YouTube. First, we pre-process the title by tokenizing the text into words and then we perform stemming using the Porter Stemmer algorithm.
Fig.~\ref{fig:groundtruth_analysis_plots_headlines} shows the top 15 stems found in titles along with their proportion for each class of videos for all the different sets of videos in our ground truth dataset.
Unsurprisingly, the top 15 stems of the Elsagate-related videos refer to popular cartoons like Peppa Pig, Mickey and Minnie mouse, Elsa, and Spiderman (see Fig.~\ref{fig:headline_normalized_mean_scores_elsagate}). 
When looking at the results, we observe that a substantial percentage of the videos that include these terms in their title are actually disturbing.
For example, from the videos that contain the terms ``spiderman'' and ``mous'', 82.6\% and 80.4\%, respectively, are disturbing.
Similar trends are observed with other terms like ``peppa'' (78.6\%), ``superhero''(76.7\%), ``pig'' (76.4\%), ``frozen'' (63.5\%), and ``elsa'' (62.5\%).
Also, we observe a small percentage of the other child-related videos that contain the terms ``felix'' (7.1\%), ``cat'' (4.2\%), and ``cartoon'' (3.8\%) are also disturbing (see Fig.~\ref{fig:headline_normalized_mean_scores_childrelated}).
These results reveal that disturbing videos on YouTube refer to seemingly ``innocent'' cartoons in their title, but in reality the content of the video is likely to be either restricted or disturbing. Note that we find these terms in suitable videos too.
This demonstrates that it is quite hard to distinguish suitable from disturbing videos by only inspecting their titles.

\noindent \textbf{Tags.} 
Tags are words that uploaders define when posting a video on YouTube. 
To study the effect of tags in this problem, we plot in Fig.~\ref{fig:groundtruth_analysis_plots_video_tags} the top 15 stems from tags found in each subset 
of videos in our ground truth dataset.
We make several observations: first, in the Elsagate-related and other child-related videos there is a substantial overlap between the stems found in the tags and title (cf. Fig.~\ref{fig:groundtruth_analysis_plots_headlines} and Fig.~\ref{fig:groundtruth_analysis_plots_video_tags}).
Second, in the Elsagate-related videos we find that suitable and disturbing classes have a considerable percentage for each tag, hence highlighting that Elsagate-related disturbing videos use the same tags as suitable videos. Inspecting these results, we find that the tags ``funni'' (47.8\%), ``elsa'' (58.7\%), ``frozen'' (57.8\%), ``cartoon'' (48.8\%), and ``anim'' (44.5\%) appear mostly in disturbing videos.
Also, ``spiderman'' (74.4\%) and ``mous'' (73.0\%) appear to have a higher portion of disturbing videos than the other tags (see Fig.~\ref{fig:video_tags_stems_normalized_mean_scores_elsagate}).
Third, we observe that the tags ``mous'' (73.0\%), ``anim'' (44.5\%), ``cartoon'' (48.8\%), ``video'' (31.5\%), ``disney'' (36.5\%), and ``kid'' (34.2\%) that appear in a considerable number of disturbing Elsagate-related videos also appear in a high portion of suitable other child-related videos (cf. Fig.~\ref{fig:video_tags_stems_normalized_mean_scores_elsagate} and Fig.~\ref{fig:video_tags_stems_normalized_mean_scores_childrelated}).
The main take-away from this analysis is that it is hard to detect disturbing content just by looking at the tags, and that popular tags are shared among disturbing and suitable videos.

\noindent \textbf{Thumbnails.} 
To study the thumbnails of the videos in our ground truth dataset, we make use of the Google Cloud Vision API\footnote{\url{https://cloud.google.com/vision/}}, which is a RESTful API that derives useful insights from images using pre-trained machine learning models. 
Using this API we are able to: 1) extract descriptive labels for all the thumbnails in our ground truth; and 2) check whether a modification was made to a thumbnail, and whether a thumbnail contains adult, medical-related, violent, and/or racy content. 
Fig.~\ref{fig:groundtruth_analysis_plots_thumbnails_stems} depicts the top 15 labels derived from the thumbnails of videos in our ground truth. 
In the Elsagate-related case, we observe that the thumbnails of disturbing videos contain similar entities as the thumbnails of both the Elsagate-related and other child-related suitable videos (cartoons, fictional characters, etc.).

Fig.~\ref{fig:groundtruth_analysis_plots_thumbnails_safesearch_categories} shows the proportion of each class for videos that contain spoofed, adult, medical-related, violent, and/or racy content.
As expected, most of the Elsagate-related videos whose thumbnails contain adult ($54.9\%$) and medical content ($53.5\%$) are restricted.
However, this is not the case for videos whose thumbnails contain spoofed ($47.4\%$), violent ($60.0\%$) or racy ($34.8\%$) content, where we observe a high number of disturbing videos (cf. Fig.~\ref{fig:thumbnails_safe_search_normalized_mean_scores_elsagate}).
Surprisingly, we notice that $100.0\%$ of the other child-related videos whose thumbnail contains violent content are suitable.
Nonetheless, after manually inspecting some of those thumbnails we notice that they depict mild cartoon violence (i.e., tom hitting jerry), which we consider as suitable.
In general, we observe that Elsagate-related videos whose thumbnail is modified with violent, racy, medical, and/or adult content are more likely to be restricted or disturbing, while this is not the case for the other child-related videos.

\noindent \textbf{Statistics.} 
Next, we examine statistics that pertain to the videos in our ground truth dataset. 
Fig.~\ref{fig:cdf_plots_views} shows the CDF of the number of views of all the videos in each distinct subset of videos in our ground truth.
We observe that Elsagate-related suitable videos have more views than disturbing videos while this is not the case for all the other types of videos.
Fig.~\ref{fig:cdf_plots_likesdislikes_fraction} shows the CDF of the fraction of likes of all the videos in each subset. 
Interestingly, we observe that in all cases disturbing and restricted videos have a higher fraction of likes compared to suitable videos, which, particularly in the case of disturbing videos, indicates manipulation to boost their ranking. 
Lastly, Fig.~\ref{fig:cdf_plots_commentsviews_fraction} shows the CDF of the fraction of comments to views.
Although, for the Elsagate-related videos the suitable and disturbing videos have a similar ratio of comments, the situation shifts when it comes to all the other types of videos where we observe a higher ratio of comments for disturbing and restricted videos compared to suitable videos.

A general take away from this ground truth analysis is that none of the videos' metadata can clearly indicate that a video is disturbing or not, thus, in most cases one (e.g., a guardian) has to carefully inspect all the available video metadata, and potentially the actual video, to accurately determine if it is safe for a toddler to watch.

\noindent \textbf{Assessing YouTube's Counter-measures.}
To assess how fast YouTube detects and removes inappropriate videos, we leverage the YouTube Data API to count the number of off-line videos (either removed by YouTube due to a Terms of Service violation or deleted by the uploader) in our manually reviewed ground truth dataset.
We note that we do not consider the videos that were already marked as age-restricted, since YouTube took the appropriate measures.

As of May 10, 2019 only 9.65\% of the suitable, 20.5\% of the disturbing, 2.5\% of the restricted, and 2.4\% of the irrelevant videos were removed, while from those that were still available, 0.0\%, 6.9\%, 1.3\%, and 0.1\%, respectively, were marked as age-restricted.
Alarmingly, the amount of the deleted disturbing and restricted videos, is considerably low.
The same stands for the amount of disturbing and restricted videos marked as age-restricted.
A potential issue here is that the videos on our dataset were recently uploaded and YouTube simply did not have time to detect them.
To test this hypothesis, we calculate the mean number of days from publication up to May, 2019: we find this hypothesis does not hold.
The mean number of days since being uploaded for the suitable, disturbing, restricted, and irrelevant videos is 866, 942, 1091, and 991, respectively, with a mean of 947 days across the entire manually reviewed ground truth dataset.
This indicate that YouTube's deployed counter-measures eliminated some of the disturbing videos, but they are unable to tackle the problem in a timely manner.

\section{Detection of Disturbing Videos} \label{sec:detectionofdisturbingvideos}

\subsection{Dataset and Feature Description}
To train and test our proposed deep learning model we use our ground truth dataset of 4,797 videos, summarized in Table~\ref{tab:final_groundtruth_dataset_details}.
For each video in our ground truth our model processes the following: 

\noindent \textbf{Title.}
Our model considers the text of the title by training an embedding layer, which encodes each word in the text in an N-dimensional vector space. 
The maximum number of words found in a title of videos in our ground truth is 21, while the size of the vocabulary is 12,023.

\noindent \textbf{Tags.} 
Similarly to the title, we encode the video tags into an N-dimensional vector space by training a separate embedding layer. 
The maximum number of tags found in a video is 78, while the size of the word vocabulary is 40,096.

\noindent \textbf{Thumbnail.} 
We scale down the thumbnail images to 299x299 while preserving all three color channels.

\noindent \textbf{Statistics.} 
We consider all available statistical metadata for videos (number of views, likes, dislikes, and comments).

\noindent \textbf{Style Features.} 
We consider some style features from the actual video (e.g., duration), the title (e.g., number of bad words), the video description (e.g., description length), and the tags (e.g., number of tags).
For this we use features proposed in~\cite{kaushal2016kidstube} that help the model to better differentiate between the videos of each class. 
Table~\ref{tab:style_features_description} summarizes the style features that we use.

\begin{table}[t!]
\centering
\resizebox{.9\columnwidth}{!}{%
\begin{tabular}{ll}
\toprule
\textbf{Type} & \textbf{Style Features Description} \\
\midrule
\textbf{Video-related} & video category, video duration \\
\textbf{Statistics-related} & ratio of \# of likes to dislikes \\
\textbf{Title-related \&} & length of title, length of description,\\
\textbf{description-related} & ratio of description to title,\\
 & jaccard sim. of title \& description, \\
 & \# '!' and '?' in title \& description, \\
 & \# emoticons in title \& description, \\
 & \# bad words in title \& description, \\
 & \# child-related words in title \& description \\
\textbf{Tags-related} & \# tags, \# bad words in tags, \\ 
 & \# child-related words in tags, \\
 & jaccard sim. of tags \& title \\
\toprule
\end{tabular}
}
\caption{List of the style features extracted from the available metadata of a video.}
\label{tab:style_features_description}
\end{table}

\subsection{Model Architecture}
\begin{figure}[t!]
\centering
\includegraphics[width=\columnwidth]{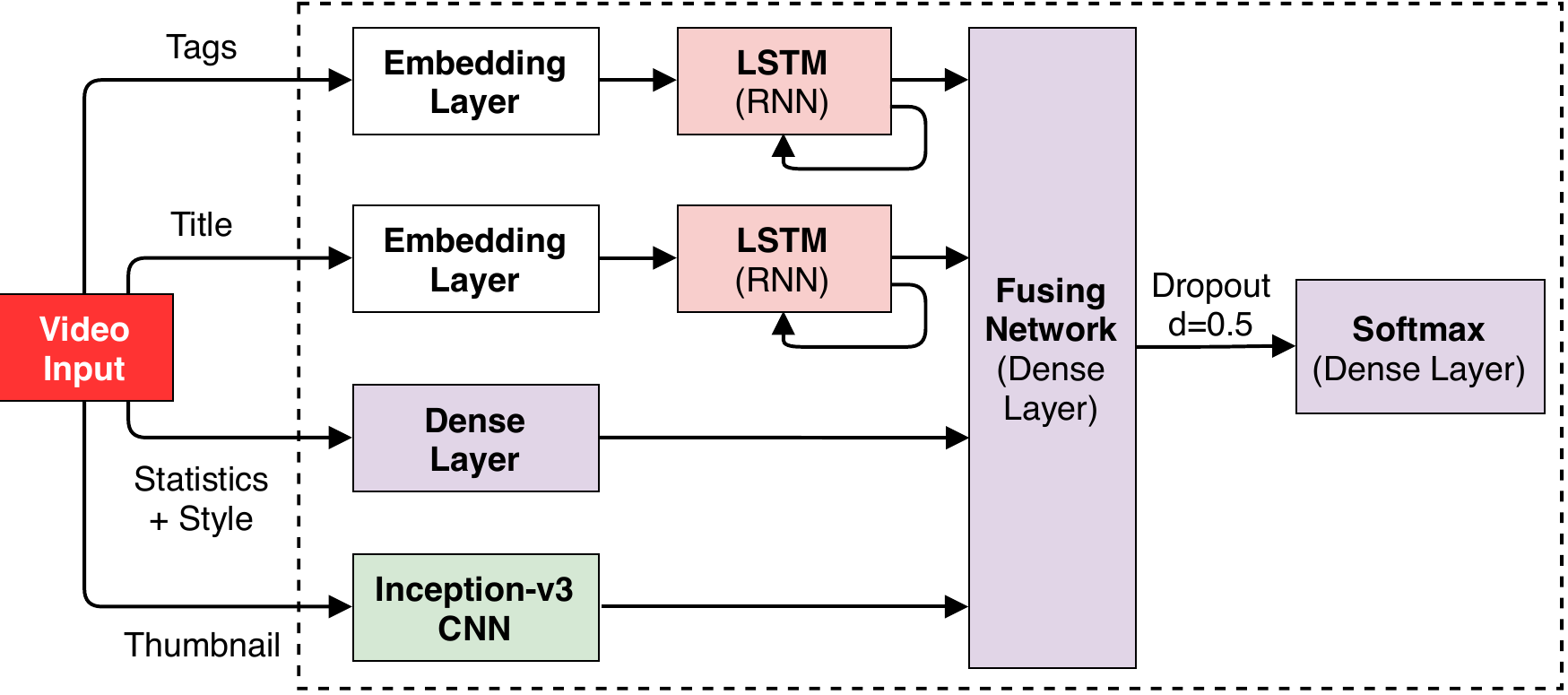}
\caption{Architecture of our deep learning model for detecting disturbing videos. The model processes almost all the video features: (a) tags; (b) title; (c) statistics \& style; and (d) thumbnail.}
\label{fig:model_architecture} 
\end{figure}

Fig.~\ref{fig:model_architecture} depicts the architecture of our classifier, which combines the above mentioned features.
Initially, the classifier consists of four different branches, where each branch processes a distinct feature type: title, tags, thumbnail, and statistics and style features.
Then the outputs of all the branches are concatenated to form a two-layer, fully connected neural network that merges their output and drives the final classification.

The title feature is fed to a trainable embedding layer that outputs a 32-dimensional vector for each word in the title text.
Then, the output of the embedding layer is fed to a Long Short-Term Memory (LSTM)~\cite{hochreiter1997long} Recurrent Neural Network (RNN) that captures the relationships between the words in the title.
For the tags, we use an architecturally identical branch trained separately from the title branch.

For thumbnails, due to the limited number of training examples in our dataset, we use transfer learning~\cite{oquab2014learning} and the pre-trained Inception-v3 Convolutional Neural Network (CNN)~\cite{Szegedy_2015_CVPR}, which is built from the large-scale ImageNet dataset.\footnote{\url{http://image-net.org/}}
We use the pre-trained CNN to extract a meaningful feature representation (2,048-dimensional vector) of each thumbnail.
Last, the statistics together with the style features are fed to a fully-connected dense neural network comprising 25 units.

The second part of our classifier is essentially a two-layer, fully-connected dense neural network. 
At the first layer, (dubbed Fusing Network), we merge together the outputs of the four branches, creating a 2,137-dimensional vector.
This vector is subsequently processed by the 512 units of the Fusing Network.
Next, to avoid possible over-fitting issues we regularize via the prominent Dropout technique~\cite{srivastava2014dropout}.
We apply a Dropout level of $d=0.5$, which means that during each iterations of training, half of the units in this layer do not update their parameters.
Finally, the output of the Fusing Network is fed to the last dense-layer neural network of four units with softmax activation, which are essentially the probabilities that a particular video is suitable, disturbing, restricted, or irrelevant.

\subsection{Experimental Evaluation} 
We implement our model using Keras~\cite{keras2015application} with TensorFlow as the backend~\cite{abadi2016tensorflow}. 
To train our model we use five-fold stratified cross-validation~\cite{arlot2010survey} and we train and test our model using all the aforementioned features. 
To deal with the data imbalance problem we use the Synthetic Minority Over-sampling technique (SMOTE)~\cite{chawla2002smote} to oversample the train set at each fold.

\begin{figure}[t!]
\centering
\includegraphics[width=2.2in]{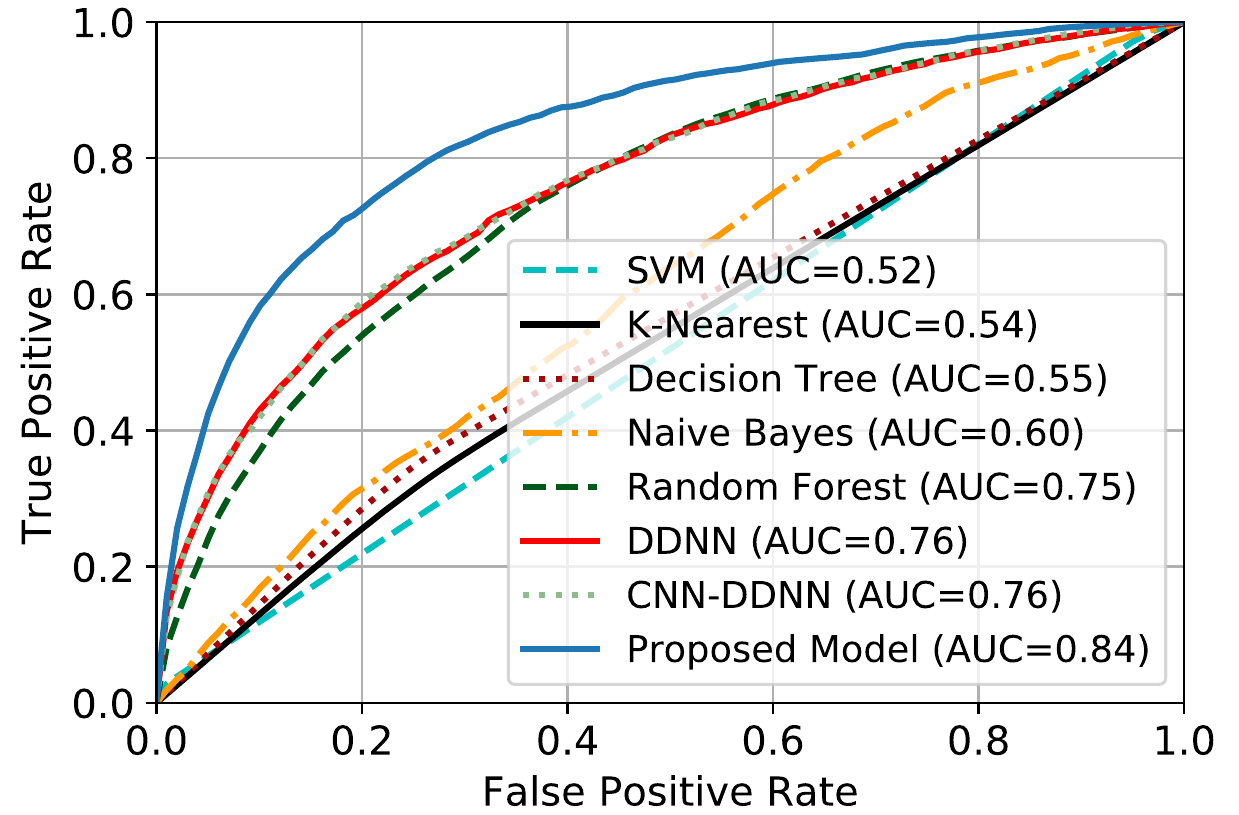}
\caption{ROC curves (and AUC) of all the baselines and of the proposed model trained for multi-class classification.}
\label{fig:all_models_multiclass_roc}
\end{figure}

For the stochastic optimization of our model, we use the Adam algorithm with an initial learning rate of $1\mathrm{e}{-5}$, and $\epsilon=1\mathrm{e}{-8}$. 
To evaluate our model, we compare it in terms of accuracy, precision, recall, F1 score, and area under the ROC curve (AUC) against the following five baselines: 
1) a Support Vector Machine (SVM) with parameters $\gamma=auto$ and $C=10.0$;
2) a K-Nearest Neighbors classifier with $n=8$ neighbors and leaf size equal to $10$; 
3) a Bernoulli Naive Bayes classifier with $a=1.0$; 
4) a Decision Tree classifier with an entropy criterion; and 
5) a Random Forest classifier with an entropy criterion and number of trees equal to $100$.
To further evaluate the performance of our model, we also compare it with two deep neural networks: 
1) a simple double dense layer network (DDNN); and
2) a CNN combined with a double dense layer network (CNN-DDNN).
For hyper-parameter tuning of all the baselines we use the grid search strategy, while for the deep neural networks we use the same hyper-parameters as with the proposed model.
For a fair comparison, we note that all the evaluated models use all the available input features.
Table~\ref{tab:performance_metrics_multiclass} reports the performance of the proposed model as well as the $7$ baselines, while Fig.~\ref{fig:all_models_multiclass_roc} shows their ROC curves.
Although the proposed model outperforms all the baselines in all performance metrics, it still has poor performance.

\begin{table}[t!]
\centering
\resizebox{.9\columnwidth}{!}{%
\begin{tabular}{lcccc}
\toprule
\textbf{Model} & \textbf{Accuracy} & \textbf{Precision} & \textbf{Recall} & \begin{tabular}[c]{@{}c@{}}\textbf{F1}\\\textbf{Score}\end{tabular}  \\
\midrule
Naive Bayes & 0.339 & 0.340 & 0.324 & 0.301 \\
K-Nearest & 0.348 & 0.301 & 0.299 & 0.297 \\
Decision Tree & 0.375 & 0.322 & 0.319 & 0.317 \\
SVM & 0.412 & 0.392 & 0.260 & 0.172 \\
Random Forest & 0.570 & 0.466 & 0.417 & 0.394 \\
DDNN & 0.467 & 0.374 & 0.368 & 0.365 \\
CNN-DDNN & 0.540 & 0.481 & 0.479 & 0.472 \\
\textbf{Proposed Model} & \textbf{0.640} & \textbf{0.495} & \textbf{0.509} & \textbf{0.478} \\
\toprule
\end{tabular}%
}
\caption{Performance metrics for the evaluated baselines and for the proposed deep learning model.}
\label{tab:performance_metrics_multiclass}
\end{table}

\begin{table}[t!]
\centering
\resizebox{\columnwidth}{!}{%
\begin{tabular}{lccccccc}
\toprule
\textbf{Input Features} & \textbf{Accuracy} & \textbf{Precision} & \textbf{Recall} & \begin{tabular}[c]{@{}c@{}}\textbf{F1}\\\textbf{Score}\end{tabular}  \\
\midrule
Thumbnail & 0.636 & 0.475 & 0.496 & 0.451 \\
Title & 0.459 & 0.352 & 0.328 & 0.301 \\
Tags & 0.395 & 0.291 & 0.304 & 0.265 \\
Style\&Stats & 0.433 & 0.348 & 0.346 & 0.288 \\
\midrule
Thumbnail, Title & 0.634 & 0.453 & 0.497 & 0.453 \\
Thumbnail, Tags & 0.629 & 0.468 & 0.493 & 0.449 \\ 
Thumbnail, Style\&Stats & 0.631 & 0.477 & 0.503 & 0.472 \\
Title, Tags & 0.485 & 0.396 & 0.376 & 0.363 \\ 
Title, Style\&Stats & 0.439 & 0.389 & 0.368 & 0.335 \\
Tags, Style\&Stats & 0.407 & 0.356 & 0.338 & 0.275 \\ 
\midrule
Title, Tags, Style\&Stats & 0.458 & 0.385 & 0.480 & 0.355 \\
Thumbnail, Tags, Style\&Stats & 0.640 & 0.479 & 0.508 & 0.477 \\ 
Thumbnail, Title, Style\&Stats & 0.630 & 0.462 & 0.501 & 0.461 \\
Thumbnail, Title, Tags & 0.636 & 0.480 & 0.509 & 0.465 \\ 
\textbf{All Input Features} & \textbf{0.640} & \textbf{0.495} & \textbf{0.509} & \textbf{0.478} \\ 
\bottomrule
\end{tabular}%
}
\caption{Performance of the proposed model trained with all the possible combinations of the four input feature types.}
\label{tab:ablation_study_details}
\end{table}

\noindent \textbf{Ablation Study.}
In an attempt to understand which of the input feature types contribute the most to the classification of disturbing videos we perform an ablation study. 
That is, we systematically remove each of the four input feature types (as well as their associated branch in the proposed model's architecture), while also training models with all the possible combinations of the four input feature types.
Again, to train and test these models we use five-fold cross validation and the oversampling technique to deal with data imbalance.
Table~\ref{tab:ablation_study_details} shows the performance metrics of all the models for each possible combination of inputs.
We observe that the thumbnail, is more important than the other input feature types for good classification performance.

\noindent \textbf{Binary Classification.}
To perform a more representative analysis of the inappropriate videos on YouTube, we need a more accurate classifier.
Thus, for the sake of our analysis in the next steps, we collapse our four labels into two general categories, by combining the suitable with the irrelevant videos into one ``appropriate'' category (3,499 videos) and the disturbing with the restricted videos into a second ``inappropriate'' category (1,348 videos).
We call the first category ``appropriate'' despite  including PG and PG-13 videos  because those videos are not aimed at toddlers (irrelevant). 
On the other hand, videos rated as PG or PG-13 that target toddlers (aged 1 to 5) are disturbing and fall under the inappropriate category. 
When such videos appear on the video recommendation list of toddlers, it is a strong indication that they are disturbing and our binary classifier
is very likely to detect them as inappropriate.

\begin{figure}[t!]
\centering
\includegraphics[width=2.2in]{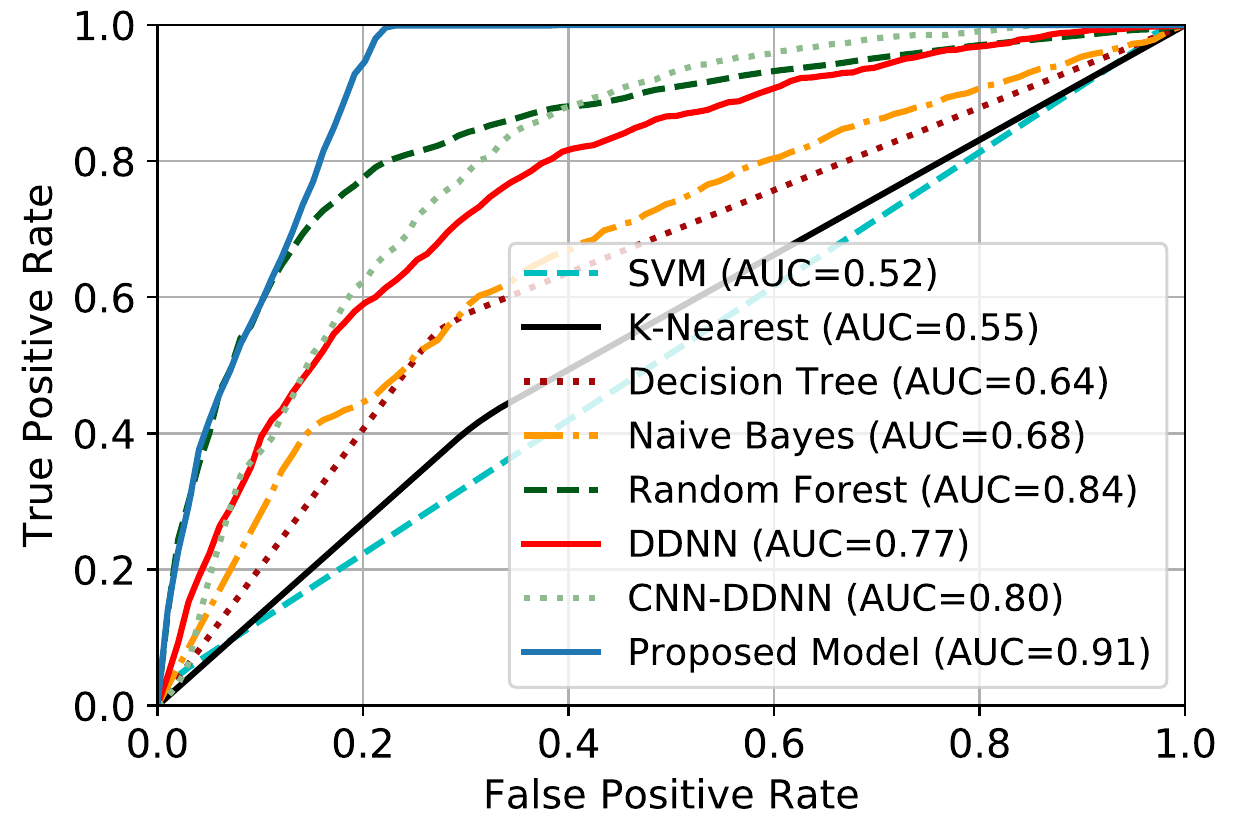}
\caption{ROC Curves of all the baselines and of the proposed model trained for binary classification.}
\label{fig:all_models_binary_roc}
\end{figure}

We train and evaluate the proposed model for binary classification on our reshaped ground truth dataset following the same approach as the one presented above.
Table~\ref{tab:performance_metrics_binary} reports the performance of our model as well as the baselines, while Fig.~\ref{fig:all_models_binary_roc} shows their ROC curves. 
We observe that our deep learning model outperforms all baseline models across all performance metrics. 
Specifically, our model substantially outperforms the CNN-DDNN model, which has the best overall performance from all the evaluated baselines, on accuracy, precision, recall, F1 score and AUC by $12.3\%$, $13.3\%$, $16.6\%$, $13.9\%$, $11.0\%$ respectively.

\begin{table}[t!]
\centering
\resizebox{.9\columnwidth}{!}{%
\begin{tabular}{lrrrr}
\toprule
\textbf{Model} & \textbf{Accuracy} & \textbf{Precision} & \textbf{Recall} & \begin{tabular}[c]{@{}c@{}}\textbf{F1}\\\textbf{Score}\end{tabular}  \\
\midrule
K-Nearest & 0.610 & 0.343 & 0.424 & 0.380 \\
Decision Tree & 0.678 & 0.442 & 0.562 & 0.495 \\
SVM & 0.718 & 0.471 & 0.029 & 0.054 \\
Naive Bayes & 0.728 & 0.522 & 0.363 & 0.428 \\
Random Forest & 0.804 & 0.740 & 0.464 & 0.569 \\
DDNN & 0.734 & 0.662 & 0.629 & 0.637 \\
CNN-DDNN & 0.720 & 0.688 & 0.724 & 0.690 \\
\textbf{Proposed Model} & \textbf{0.843} & \textbf{0.821} & \textbf{0.890} & \textbf{0.829} \\
\toprule
\end{tabular}%
}
\caption{Performance of the evaluated baselines trained for binary classification and of our proposed binary classifier.}
\label{tab:performance_metrics_binary}
\end{table}

\section{Analysis} \label{sec:analysis}
In this section, we study the interplay of appropriate and inappropriate videos on YouTube using our binary classifier.
First, we assess the prevalence of inappropriate videos in each subset of videos in our dataset and investigate how likely it is for YouTube to recommend an inappropriate video.
Second, we perform live random walks on YouTube's recommendation graph to simulate the behavior of a toddler that selects videos based on the recommendations.

\begin{table}[t!]
\centering
\resizebox{0.9\columnwidth}{!}{%
\begin{tabular}{lrr}
\toprule
\textbf{Videos subset} & \textbf{Appropriate (\%)} & \textbf{Inappropriate (\%)} \\ 
\midrule
Elsagate-related & 230,890 (98.95\%) & 2,447 (1.05\%) \\
Other Child-related & 154,262 (99.55\%) & 695 (0.45\%) \\
Random & 478,420 (99.28\%) & 3,487 (0.72\%) \\
Popular & 10,947 (99.75\%) & 27 (0.25\%) \\
\toprule
\end{tabular}%
}
\caption{Number of appropriate and inappropriate videos found in each subset of videos in our dataset.}
\label{tab:appropriate_inappropriate_videos_in_dataset}
\end{table}

\begin{table*}[t!]
\centering
\resizebox{0.85\textwidth}{!}{%
\begin{tabular}{llrrrr}
\toprule
\textbf{Source} & \textbf{Destination} & \textbf{Elsagate-related (\%)} & \textbf{Other Child-related (\%)} & \textbf{Random (\%)} & \textbf{Popular (\%)} \\ 
\midrule
Appropriate 	& Appropriate 	& 917,319 (97.80\%) & 648,406 (99.49\%) & 1,319,518 (98.82\%) & 34,764 (99.12\%) \\
Appropriate 	& Inappropriate 	& 5,951 (0.64\%) 	& 1,681 (0.26\%) 	& 7,014 (0.53\%)  & 64 (0.18\%) \\
Inappropriate 	& Appropriate 	& 14,202 (1.51\%) 	& 1,542 (0.24\%) 	& 7,946 (0.59\%) 	& 246 (0.70\%) \\
Inappropriate 	& Inappropriate 	& 478 (0.05\%) 		& 72 (0.01\%) 	& 831 (0.06\%) 	& 0 (0.00\%) \\ 
\toprule
\end{tabular}%
}
\caption{Number of transitions between appropriate and inappropriate videos for each subset of videos in our dataset.}
\label{tab:graph_transitions_in_dataset_videos}
\end{table*}

\subsection{Recommendation Graph Analysis}
First, we investigate the prevalence of inappropriate videos in each subset of videos in our dataset by running our binary classifier on the whole dataset, which allows us to find which videos are inappropriate or appropriate.
Table~\ref{tab:appropriate_inappropriate_videos_in_dataset} shows the number of appropriate and inappropriate videos found in each subset.
For the Elsagate-related videos, we find 231K (98.9\%) appropriate videos and 2.5K (1.1\%) inappropriate videos, while the proportion of inappropriate videos is a bit lower in the set of other child-related videos (0.4\% inappropriate and 99.5\% appropriate).
These findings highlight the gravity of the problem: a parent searching on YouTube with simple toddler-oriented keywords and casually selecting from the recommended videos, is likely to expose their child to inappropriate videos.

But what is the interplay between the inappropriate and appropriate videos in each subset?
To shed light to this question, we create a directed graph for each subset of videos, where nodes are videos, and edges are recommended videos (up to 10 videos due to our data collection methodology).
For instance, if $video_2$ is recommended via $video_1$ then we add an edge from $video_1$ to $video_2$.
Then, for each video in each graph, we calculate the out-degree in terms of appropriate and inappropriate labeled nodes.
From here, we can count the number of \emph{transitions} the graph makes between differently labeled nodes.
Table~\ref{tab:graph_transitions_in_dataset_videos} summarizes the percentages of transitions between the two classes of videos in each subset.
Unsurprisingly, we find that most of the transitions in each subset (98\%-99\%), are between appropriate videos, which is mainly because of the large number of appropriate videos in each set.
We also find that when a toddler watches an Elsagate-related benign video, if she randomly follows one of the top ten recommended videos, there is a $0.6\%$ probability that she will end up at a disturbing or restricted video.
Taken altogether, these findings show that the problem of toddler-oriented inappropriate videos on YouTube is notable, especially when considering YouTube's massive scale and the large number of views of toddler-oriented videos.That is, there is a non-negligible chance that a toddler will be recommended an inappropriate video when watching an appropriate video.

\begin{figure}[t!]
\centering
\subfigure[]{\includegraphics[width=.49\columnwidth]{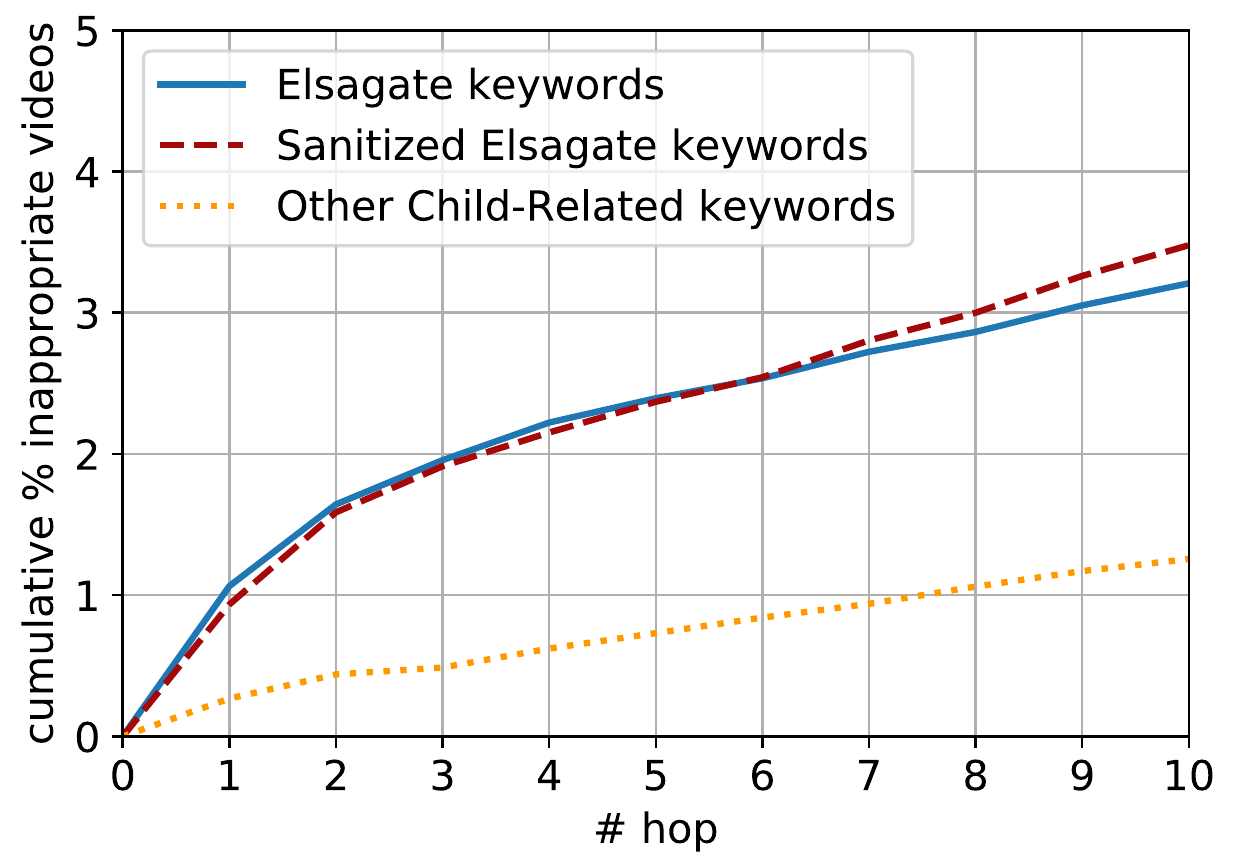}\label{fig:cumulative_percentage_disturbing_in_hops_classes}}
\subfigure[]{\includegraphics[width=.49\columnwidth]{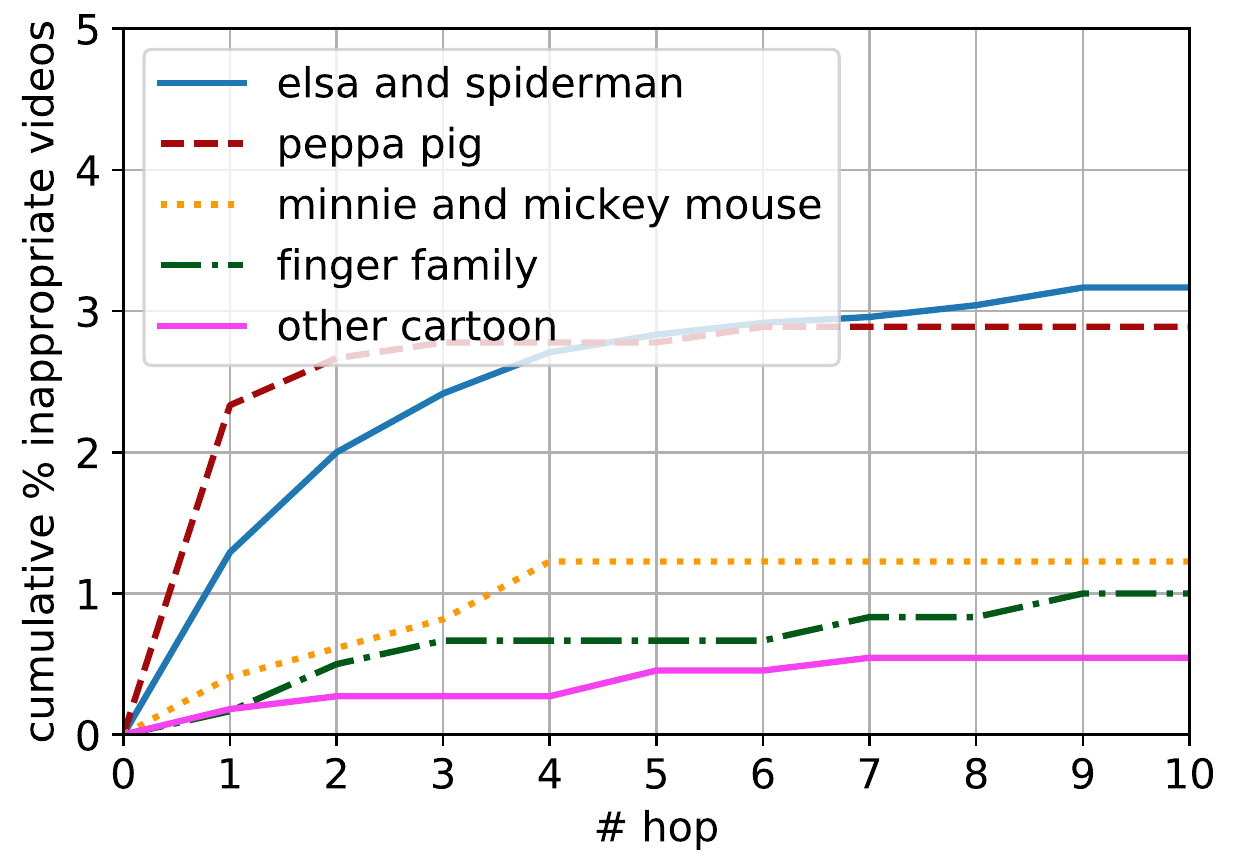}\label{fig:cumulative_percentage_disturbing_in_hops_clusters}}
\caption{Cumulative percentage of inappropriate videos encountered at each hop for: (a) Elsagate-related, sanitized Elsagate-related, and other child-related seed keywords; and (b) clusters of seed keywords.}
\label{fig:all_random_walks_plots}
\end{figure}

\subsection{How likely is it for a toddler to come across inappropriate videos?}
In the previous section, we showed that the problem of toddler-oriented videos is prevalent enough to be cause for concern. 
However, it is unclear whether the previous results generalize to YouTube at large since our dataset is based on a snowball sampling up to three hops from a set of seed videos. In reality though, YouTube comprises billions of videos, which are recommended over many hops within YouTube's recommendation graph.
Therefore, to assess how prominent the problem is on a larger scale, we perform live random walks on YouTube's recommendation graph.
This allow us to simulate the behavior of a ``random toddler'' who searches the platform for a video and then he watches several videos according to the recommendations.
To do this, we use the lists of Elsagate-related and other child-related seed keywords used for constructing our dataset, as well as a list of sanitized Elsagate-related seed keywords which we construct by stripping all the inappropriate words from all the Elsagate-related keywords using a dictionary of inappropriate words\footnote{\url{https://tinyurl.com/yxb4kmxg}}.
We do this to assess the degree of the problem around Elsagate-related videos while ensuring that we are not biasing the search results with any sensitive words.

For each seed keyword, we initially perform a search query on YouTube and randomly select one video from the top ten results.
Then, we obtain the recommendations of the video and select one randomly. 
We iterate with the same process until we reach ten hops, which constitutes the end of a single random walk.
We repeat this operation for 100 random walks for each seed keyword, while at the same time classifying each video we visit, using our binary classifier.

First, we group the random walks based on the keywords used to seed them.
Fig.~\ref{fig:cumulative_percentage_disturbing_in_hops_classes} shows the cumulative percentage of inappropriate videos encountered at each hop of the random walks for Elsagate-related, sanitized Elsagate-related, and other child-related search keywords.
We observe that, when using sanitized Elsagate-related keywords, we find at least one inappropriate video in $3.5\%$ of the walks, while for the other child-related keywords we find at least one inappropriate video in $1.3\%$ of the walks.
We also observe that most of the inappropriate videos are found early in our random walks (i.e., at the first hop) and this number decreases as the number of hops increases.
These findings highlight that the problem of inappropriate videos on YouTube emerges quite early when users are browsing the platform starting from benign toddler-oriented search terms.

Next, to assess whether our results change according to the content of the videos we use the k-means clustering algorithm~\cite{hartigan1979algorithm} to create clusters from all the seed keywords.
Then, we manually inspect the clusters and associate a label to each cluster.
We create five clusters: 1) ``Elsa and Spiderman'' (24 keywords); 2) ``Peppa Pig'' (9 keywords); 3) ``Finger Family'' (6 keywords); 4) ``Minnie and Mickey mouse'' (5 keywords); and 5) ``Other Cartoon'' (11 keywords).
Then, based on the clusters we group the random walks.
Fig.~\ref{fig:cumulative_percentage_disturbing_in_hops_clusters} shows the cumulative percentage of inappropriate videos encountered at each hop for each cluster.
We observe interesting differences across the clusters: specifically, we observe the higher percentages in the ``Elsa and Spiderman'' (3.2\%), and ``Peppa pig'' (2.9\%) cluster, whereas for the clusters ``finger family'' (1.0\%) and ``other cartoon'' (0.5\%) we observe a lower percentage of walks with inappropriate videos.
Also, we find that most of the inappropriate videos are found in the beginning of the random walks in particular for the clusters ``Peppa pig'' (2.3\%) and ``Elsa and Spiderman'' (1.3\%) (see first hop in Fig.~\ref{fig:cumulative_percentage_disturbing_in_hops_clusters}).

Note that, by merging the two classes in the binary classifier while seeking out disturbing videos with short random walks from suitable videos, we correct for the misclassification of disturbing videos as restricted. 
That is, an NC-17 video in the proximity of benign toddler-oriented videos could be erroneously classified as restricted by the multi-class classifier (because of similarities with the NC-17 videos that do not target toddlers).
However, due to this proximity, this is probably an NC-17 video that targets toddlers and should have therefore been classified as disturbing. 
Thus, the vast majority of inappropriate videos detected during the random walks are expected to be disturbing. 
In fact, 84.6\% of the detected inappropriate videos are disturbing (obtained by inspecting all the 338 detected inappropriate videos).
On the other hand, videos that would be classified as irrelevant by the multi-class classifier, fall under the appropriate category of the binary classifier. 
However, the training set for the appropriate category includes irrelevant videos, which include PG and PG-13 videos that do not target toddlers. 
Therefore, the binary classifier may classify such a video that is in the proximity of suitable videos as appropriate.
However, a PG and PG-13 video in the proximity of suitable videos is likely to actually be disturbing, thus inappropriate. 
This negatively affects the accuracy of the binary classifier. 
Yet, only 1.6\% of the videos encountered during the random walks and classified as appropriate were in fact disturbing (obtained by sampling 300 of the 176,619 detected appropriate videos).

\section{Related Work} \label{sec:relatedwork}
Prior work studied YouTube videos with inappropriate content for children, as well as spam, hate or malicious activity.

\noindent \textbf{Inappropriate Content for Children.}
Several studies focused on understanding videos that target young children, and how they interact with them and the platform.
\cite{buzzi2011children} suggests the addition of extra parental controls on YouTube to prevent children from accessing inappropriate content.
\cite{araujo2017characterizing} study the audience profiles and comments posted on YouTube videos in popular children-oriented channels, and conclude that children under the age of 13 use YouTube and are exposed to advertising, inappropriate content, and privacy issues.
\cite{eickhoff2010identifying} propose a binary classifier, based on video metadata, for identifying suitable YouTube videos for children.
\cite{kaushal2016kidstube} focus on the characterization and detection of unsafe content for children and its promoters on YouTube.
They propose a machine learning classifier that considers a set of video-, user-, and comment-level features for the detection of users that promote unsafe content. 
\cite{ishikawa2019combating} studied the Elsagate phenomenon and they propose a deep learning model for detecting Elsagate content on YouTube trained on a unannotated dataset of 3K videos.
\cite{singh2019kidsguard} focus on the detection of child unsafe content. 
They propose an LSTM-based deep neural network for the detection of unsafe content.
Deep neural network based architectures have also been proposed for large-scale video recommendation on YouTube~\cite{covington2016deep}.

\noindent \textbf{Spam, Hate and other Malicious Activity.}
A large body of previous work focused on the detection of malicious activity on YouTube.
\cite{sureka2010mining} use social network analysis techniques to discover hate and extremist YouTube videos, as well as hidden communities in the ecosystem. 
\cite{agarwal2014focused} develop a binary classifier trained with user and video features for detecting YouTube videos that promote hate and extremism.
\cite{giannakopoulos2010multimodal} use video, audio, and textual features to train classifier for detecting violent YouTube videos.
\cite{ottoni2018analyzing} perform an in-depth analysis on video comments posted by alt-right channels on YouTube.
They conclude that the comments of a video are a better indicator for detecting alt-right videos when compared to the video's title.
\cite{aggarwal2014mining} use video features for detecting videos violating privacy or promoting harassment.
With regard to spam detection,~\cite{chowdury2013data} explore video attributes that may enable the detection of spam videos on YouTube.
\cite{sureka2011mining} focuses on both user features and comment activity logs to propose formulas/rules that can accurately detect spamming YouTube users.
Using similar features, ~\cite{bulakh2014identifying} characterize and identify fraudulently promoted YouTube videos.
\cite{chaudhary2013contextual} use only video features, and propose a one-class classifier approach for detecting spam videos.

\cite{benevenuto2012practical} propose two supervised classification algorithms to detect spammers, promoters, and legitimate YouTube users.
Also, in an effort to improve the performance of spam filtering on the platform,~\cite{alberto2015tubespam} test numerous approaches and propose a tool, based on Naive Bayes, that filters spam comments on YouTube.
Finally,~\cite{zannettou2018good} propose a classifier for identifying videos that use manipulative techniques to increase their views (i.e., clickbait).

In contrast to all the above inappropriate content and malicious activity studies, we are the first to focus on the characterization and detection of \emph{disturbing videos}, i.e., inappropriate videos that explicitly target toddlers.
We collect thousands of YouTube videos and manually annotate them according to four relevant categories.
We develop a deep learning classifier that can detect inappropriate videos with an accuracy of $84.3\%$.
By classifying and analyzing these videos, we shed light on the prevalence of the problem on YouTube, and how likely it is for an inappropriate video to be served to a toddler who casually browses the platform.

\section{Conclusion and Discussion} \label{sec:conclusions}
An increasing number of young children are shifting from broadcast to streaming video consumption, with YouTube providing an endless array of content tailored toward young viewers.
While much of this content is age-appropriate, there is also an alarming amount of inappropriate material available.
In this paper, we present the first characterization of inappropriate or disturbing videos targeted at toddlers.
From a ground truth labeled dataset, we develop a deep learning classifier that achieves an accuracy of $84.3\%$.
We leverage this classifier to perform a large-scale study of toddler-oriented content on YouTube, finding $1.05\%$ of the 233,337 Elsagate-related videos in our dataset to be inappropriate. 
Even worse, we discover a $3.5\%$ chance of a toddler who starts by watching appropriate videos to be recommended inappropriate ones within ten recommendations.

Although scientific debate (and public opinion) on the risks associated with ``screen time'' for young children is still on going, based on our 
findings, we believe a more pressing concern to be the dangers of crowd-sourced, uncurated content combined with engagement oriented, gameable recommendation systems.
Considering the advent of algorithmic content creation (e.g., ``deep fakes'') %
and the monetization opportunities on sites like YouTube, there is no reason to believe there will be an organic end to this problem.
Our classifier, and the insights gained from our analysis, can be used as a starting point to gain a deeper understanding and begin mitigating this issue.

Note that in this work, we collect and analyze a large number of Elsagate-related, other child-related, as well as random and popular videos available on YouTube.
Although not representative of the entirety of YouTube, we believe that the set of seed keywords (Elsagate-related and other child-related keywords) cover a wide range of child-related content available on the platform.
With regard to our sampling process, we believe that by including a wide range of child-related content as well as other types of videos in our ground truth dataset, we aid the proposed model to generalize to different types of videos that are available on YouTube.

\noindent \textbf{Limitations.} 
Finally, we discuss some of the limitations of this work.
First, we collect and analyze videos only from YouTube and not YouTube Kids. 
This is because YouTube does not provide an open API for collecting videos that appear on YouTube Kids. 
However, according to YouTube, only videos marked as age-restricted are excluded from YouTube Kids unless specific settings are set by the parent\footnote{\url{https://tinyurl.com/youtube-kids-restricted}}.
Second, we acknowledge that the performance of our classifier is highly affected by the small training size: we were unable to provide a larger annotated dataset mainly due to lack of resources for the annotation process.

\noindent \textbf{Acknowledgments.}
This project has received funding from the European Union’s Horizon 2020 Research and Innovation program under the Marie Skłodowska-Curie ENCASE project (Grant Agreement No. 691025) and from the National Science Foundation under grant CNS-1942610.

\small
\bibliographystyle{abbrv}

\end{document}